\documentclass[aps,floats]{revtex4}
\usepackage{amsmath,amssymb}
\usepackage{graphicx,epsfig}

\begin{document}
\bibliographystyle {plain}

\def\oppropto{\mathop{\propto}} 
\def\opsimeq{\mathop{\simeq}}
\def\opoverderline{\mathop{\overline}}
\def\operarrow{\mathop{\longrightarrow}}
\def\opsim{\mathop{\sim}} 
\def\opmin{\mathop{\min}} 
\def\opmax{\mathop{\max}} 

\def\fig#1#2{\includegraphics[height=#1]{#2}}
\def\figx#1#2{\includegraphics[width=#1]{#2}}


\title{ Many Body Localization Transition in the strong disorder limit : \\
entanglement entropy from the statistics of rare extensive resonances  }


 \author{ C\'ecile Monthus }
  \affiliation{ Institut de Physique Th\'{e}orique, 
Universit\'e Paris Saclay, CNRS, CEA,
 91191 Gif-sur-Yvette, France}

\begin{abstract}

The space of one-dimensional disordered interacting quantum models displaying a Many-Body-Localization Transition seems sufficiently rich to produce critical points with level statistics interpolating continuously between the Poisson statistics of the Localized phase and the Wigner-Dyson statistics of the Delocalized Phase. In this paper, we consider the strong disorder limit of the MBL transition, where the level statistics at the MBL critical point is close to the Poisson statistics. We analyse a one-dimensional quantum spin model, in order to determine the statistical properties of the rare extensive resonances that are needed to destabilize the MBL phase. At criticality, we find that the entanglement entropy can grow with an exponent $0<\alpha < 1$ anywhere between the area law $\alpha=0$ and the volume law $\alpha=1$, as a function of the resonances properties, while the entanglement spectrum follows the strong multifractality statistics. In the MBL phase near criticality, we obtain the simple value $\nu=1$ for the correlation length exponent. Independently of the strong disorder limit, we explain why for the Many-Body-Localization transition concerning individual eigenstates, the correlation length exponent $\nu$ is not constrained by the usual Harris inequality $\nu \geq 2/d$, so that there is no theoretical inconsistency with the best numerical measure $\nu = 0.8 (3)$ obtained by D. J. Luitz, N. Laflorencie and F. Alet, Phys. Rev. B 91, 081103 (2015).

\end{abstract}

\maketitle

\section{ Introduction }

The thermalization of isolated many-body quantum systems is nowadays discussed in terms of the Eigenstate Thermalization Hypothesis (E.T.H.) \cite{deutsch,srednicki,nature,mite}
 (see the review \cite{rigol} and references therein) : the idea is that each many-body excited eigenstate $\vert \psi > $
is 'thermal', i.e. the reduced density matrix $\rho_A$ of a sub-region $A$ 
corresponds the thermal density matrix
\begin{eqnarray}
\rho_{A}^{(\beta)} = \frac{ e^{- \beta H_A} }{ Tr_A (e^{- \beta H_A}) }
\label{rhoAthermal}
\end{eqnarray}
where the inverse temperature $\beta$ selects the correct average energy
corresponding to the initial quantum state $\vert \psi > $.
Then the Von Neumann entanglement entropy of the region $A$ with the complementary region
 \begin{eqnarray}
S_A \equiv - Tr_A(\rho_{A} \ln \rho_{A})
\label{saent}
\end{eqnarray}
coincides with the thermal entropy, which is extensive with respect to the volume $L_A^d$ of the region $A$
 \begin{eqnarray}
S_A \opsimeq_{ETH}  - Tr_A(\rho^{th}_{A} \ln \rho^{th}_{A})  = s^{th}(\beta) L_A^d
\label{sthermal}
\end{eqnarray}

However the presence of strong disorder can prevent this thermalization and lead to the phenomenon of
Many-Body-Localization  (MBL) (see the recent reviews \cite{revue_huse,revue_altman} and references therein),
where the disorder-averaged entanglement entropy of Eq. \ref{saent} follows instead the area-law \cite{bauer}
\begin{eqnarray}
\overline{ S_A } \oppropto_{MBL}    L_A^{d-1}
\label{area}
\end{eqnarray}
So in the MBL phase, excited states are somewhat similar to ground states,
with efficient representation via Density-Matrix-RG or Matrix Product States \cite{pekker1,pekker2,friesdorf,sondhi}
 and Tensor Networks \cite{tensor}. The Fisher Strong Disorder Real Space RG
  to construct the ground states of random quantum spin models \cite{fisher_AF,fisher,fisherreview} has been extended into the Strong Disorder RG procedure for the unitary dynamics \cite{vosk_dyn1,vosk_dyn2},
and into the RSRG-X procedure in order to construct the whole set of excited eigenstates 
 \cite{rsrgx,rsrgx_moore,vasseur_rsrgx,yang_rsrgx,rsrgx_bifurcation,c_emergent}.
This construction is actually possible only because the MBL phase
is characterized by an extensive number 
of emergent localized conserved operators
\cite{emergent_swingle,emergent_serbyn,emergent_huse,emergent_ent,imbrie,serbyn_quench,emergent_vidal,emergent_ros},
but breaks down as the MBL transition towards delocalization is approached.
As a consequence, the current RG descriptions of the MBL transition 
 are based on different types of RG rules 
concerning either the entanglement \cite {vosk_rgentanglement}
 or the resonances \cite{vasseur_resonant}.

From the point of view of the entanglement entropy of Eq. \ref{saent},
a natural question is whether at criticality, it is possible to
obtain an entanglement power-law growth  
\begin{eqnarray}
\overline{ S_A } \opsimeq_{criti}    L_A^{\alpha} 
\label{criti}
\end{eqnarray}
intermediate between the area-law and the volume law $d-1\leq \alpha \leq d$,
or with possibly some logarithmic corrections.
Let us now concentrate on the one-dimensional case $d=1$.
If one assumes some standard finite-size-scaling form in the critical region
 in terms of the diverging correlation length $\xi$ 
\begin{eqnarray}
\overline{ S_A } \opsimeq_{FSS}    L_A^{\alpha} \ \  \Phi \left( \frac{L_A}{\xi} \right)
\label{fss}
\end{eqnarray}
one obtains that the matching with the area-law of the MBL phase (Eq. \ref{area}
in $d=1$) requires the divergence as
\begin{eqnarray}
\overline{ S_A } \oppropto_{critical,loc}   \xi^{\alpha} 
\label{fssloc}
\end{eqnarray}
whereas the matching with the volume-law of the delocalized phase in $d=1$
requires a vanishing coefficient of the volume-law
\begin{eqnarray}
\overline{ S_A } \oppropto_{critical,deloc} 
   L_A \left( \frac{1}{\xi} \right)^{1-\alpha}
\label{fssdeloc}
\end{eqnarray}
As discussed in detail in \cite{grover,harrisMBL}, 
one should then distinguish two possibilities :

(i) if the transition is directly towards the thermal ergodic phase satisfying E.T.H., 
the continuous vanishing of the coefficient $\left( \frac{1}{\xi} \right)^{1-\alpha} $ of the volume-law of Eq. \ref{fssdeloc}
is actually forbidden by the strong-subadditivity property \cite{grover},
and the only possibility is that the critical point is itself thermal,
i.e. it should satisfy the volume law $\alpha=1$ 
with the finite thermal coefficient given by Eq. \ref{sthermal}.
This scenario was found numerically \cite{kjall,alet} and via the RG based on entanglement \cite {vosk_rgentanglement}
 or resonances \cite{vasseur_resonant}. Then the difference between the critical point and the delocalized phase
is not visible in the disorder-averaged entanglement entropy, but in its variance \cite{kjall,alet,vosk_rgentanglement}
and in the dynamical properties \cite {vosk_rgentanglement,vasseur_resonant}.

(ii) if the transition is towards a delocalized non-ergodic phase,
i.e. a phase satisfying the volume law, 
but not the E.T.H. with the thermal coefficient fixed by Eq. \ref{sthermal},
then the continuous vanishing of the coefficient of the volume-law
 of Eq. \ref{fssdeloc}
is possible, and the exponent $\alpha$ is not a priori fixed by the strong-subadditivity property (see \cite{grover} for more details).
This scenario is suggested by the point of view that the MBL transition is
 somewhat similar
to an Anderson Localization transition
 in the Hilbert space of 'infinite dimensionality'
 as a consequence of the exponential growth of the size of the Hilbert space 
with the volume \cite{levitov,gornyi_fock,vadim,us_mblaoki,luca,gornyi,c_mblstrongmultif}.

In this paper, we consider the strong disorder
 limit of a one-dimensional quantum spin model
in order to analyze the statistical properties of resonances that are needed to destabilize an MBL eigenstate satisfying the area-law $\overline{ S_A } =O(1)$ and to determine the entanglement properties of the obtained critical state.
The paper is organized as follows.
In section \ref{sec_perturb}, we introduce
 the one-dimensional quantum spin model and derive the entanglement spectrum
in terms of the couplings.
In section \ref{sec_ent}, the statistical properties of the entanglement spectrum
are studied in terms of L\'evy variables. The behavior of the entanglement entropy
is analyzed in Section \ref{sec_omega}.
Finally in section \ref{sec_multif}, the 
multifractal statistics of the entanglement spectrum is obtained.
 Our conclusions are summarized in section \ref{sec_conclusion}.
 Two appendices contain complementary computations.

\section{ Singular Perturbation from the Many-Body Localized phase }

\label{sec_perturb}

Before we introduce the MBL model, we need to explain the motivations 
coming from strong disorder limit of the Anderson Localization Transition.

\subsection{ Motivation :  strong disorder limit of the Anderson Localization Transition }
 
\label{anderson}

At Anderson localization transitions, 
the critical states can be more or less multifractal 
(see the review \cite{mirlinrevue} and references therein) :

(i) for the short-ranged tight-binding model in dimension $d$, 
there is a continuous interpolation 
between the 'weak multifractality' regime in $d=2+\epsilon$ and the 'strong multifractality' in high dimension $d$.

(ii) for the tight-binding model with long-ranged hopping
\begin{eqnarray}
V(r) =  \frac{ V}{r^a} 
\label{prbm}
\end{eqnarray}
criticality is known to occur exactly at $a_c=d$,
as can be understood from the scaling of the difference between energy levels 
\begin{eqnarray}
\Delta E(L^d)  \propto L^{-d}
\label{levelspacinganderson}
\end{eqnarray}
Here there is also a continuous family of critical points 
as a function of the amplitude $V$, that interpolates 
 between the 'weak multifractality' regime for large $V \to +\infty$ 
and the 'strong multifractality' regime for small $V \to 0$ \cite{mirlinrevue}.

From the point of view of the statistics of the energy levels,
the 'weak multifractality' of wavefunctions corresponds
to a level statistics close to the Wigner Dyson statistics
 of the delocalized phase,
with a strong level repulsion and a vanishing level compressibility $\chi \to 0$,
whereas the 'strong multifractality' of wavefunctions corresponds
to a level statistics close to the Poisson statistics of the localized phase,
with a very weak level repulsion and a level compressibility
close to unity $\chi \simeq 1 $ (see more details in \cite{mirlinrevue}).

The 'strong multifractality'
regime $ V \to 0$ has been analyzed via the 
Levitov renormalization method that takes into account the
resonances occuring at various scales \cite{levitov1,levitov2,levitov3,levitov4,mirlin_evers,fyodorov,fyodorovrigorous}
or other methods \cite{oleg1,oleg2,oleg3,oleg4,olivier_per,olivier_strong,olivier_conjecture}.

In a previous work \cite{us_strongmultif}, we
 have described how these strong multifractality results
for small aamplitude $V$ in the Long-Ranged-Hopping model 
can be reproduced via the first-order perturbation theory
with respect to the completely localized basis, provided one takes into account
the broad L\'evy statistics of the terms involved in the perturbative expansion.

\subsection{ Model and notations } 

\label{sec_toy}

In the present paper, we wish to adapt
 the above strategy for the MBL transition as follows :
 instead of the standard nearest-neighbor MBL models,
we wish to consider a {\it toy model } where the Hamiltonian contains
 direct couplings between any pair of configurations in the Hilbert space,
with a vanishing small amplitude and with an appropriate decay as a function of
the distance in Hilbert space in order to reach an MBL critical point.
This critical point will be then 
as close as possible to the Many-Body-Localized Phase, and
 can be studied via singular perturbation theory 
around the completely localized limit.

So we consider a one-dimensional model involving $L$ quantum spins 
\begin{eqnarray}
H && =H_0+ V 
\label{Htot}
\end{eqnarray}
with the following properties.
The Hamiltonian $H_0$ is chosen as the simplest possible 
Many-Body-Localized Hamiltonian
\begin{eqnarray}
H_0 && = - \sum_{i=1}^{L} h_i \sigma_i^z
\label{H0}
\end{eqnarray}
where the fields $h_i$ are random variables drawn with the Gaussian distribution
of variance $W^2$
\begin{eqnarray}
G_W(h) = \frac{1}{\sqrt { 2  \pi W^2 }} e^{- \frac{ h^2}{2  W^2} }
\label{gauss}
\end{eqnarray}
The $2^{L}$ eigenstates are simply given by the tensor products
\begin{eqnarray}
\vert \psi^{(0)}_{S_1,..,S_{L}} > && \equiv \vert S_1 > \otimes \vert S_2 > ... \otimes \vert S_{L} >
\label{psizero}
\end{eqnarray}
with the random energies
\begin{eqnarray}
E^{(0)}_{S_1,..,S_{L}} =- \sum_{i=1}^{L} h_i S_i
\label{e0}
\end{eqnarray}
For instance, the ground state corresponds to the choice $S_i={\rm sgn} (h_i)$
and has the extensive energy
\begin{eqnarray}
E^{(0)}_{GS} = - \sum_{i=1}^{L} \vert h_i \vert
\label{egs}
\end{eqnarray}
In the following, we focus on the middle of the spectrum, 
where the density of states follows the Gaussian
of zero mean and of variance $(LW^2)$
\begin{eqnarray}
\rho_0(E) = \frac{1}{\sqrt { 2  \pi L W^2 }} e^{- \frac{ E^2}{2  L W^2} }
\label{gaussrho}
\end{eqnarray}
Since there are $2^L$ levels, the level spacing near zero energy scales as
\begin{eqnarray}
\Delta E(L)  \propto L^{\frac{1}{2}} 2^{-L}
\label{levelspacing}
\end{eqnarray}

In analogy with Eq. \ref{prbm},
we wish to introduce a small perturbation $V$ that produces a direct coupling to all other $(2^L-1)$ states of the Hilbert space
\begin{eqnarray}
V && = - \sum_{k=1}^{L}  \sum_{1 \leq i_1< i_2..<i_k \leq {L}} 
J_{i_1,..,i_k}  \sigma^x_{i_1} \sigma^x_{i_2} ... \sigma^x_{i_k}
\label{Vperturbation}
\end{eqnarray}
The couplings $ J_{i_1,..,i_k} $ are assumed to be of small amplitude, 
but they should be able
 to produce resonances at all scales. 
Their scaling should thus be directly related
to the level spacing of Eq. \ref{levelspacing}, and in particular they should decay exponentially in space.
However the precise conditions will be discussed later, 
and it is clearer to write the perturbation theory
in the arbitrary small couplings $ J_{i_1,..,i_k}$,
first for the eigenstates, then for the corresponding reduced density matrices,
and finally for the entanglement spectrum.

 \subsection{ First order perturbation theory for the eigenstates}

At first order in the perturbation $V$, the eigenvalues of Eq. \ref{e0} are unchanged 
 \begin{eqnarray}
E^{(1)}_{S_1,..,S_{L}} = E^{(0)}_{S_1,..,S_{L}} + < \psi^{(0)}_{S_1,..,S_{L}} \vert V \vert \psi^{(0)}_{S_1,..,S_{L}} > = E^{(0)}_{S_1,..,S_{L}}
\label{e1}
\end{eqnarray}
and the eigenstates read
\begin{eqnarray}
 \vert \psi^{(1)}_{S_1,..,S_{L}} > 
&& =  \vert \psi^{(0)}_{S_1,..,S_{L}} >
+ \sum_{ \{S_i' \} } \vert \psi^{(0)}_{S_1',..,S_{L}'} > \frac{ < \psi^{(0)}_{S_1',..,S_{L}'} \vert V\vert \psi^{(0)}_{S_1,..,S_{L}} > }{ E^{(0)}_{S_1,..,S_{L}} -  E^{(0)}_{S_1',..,S_{L}'}}
\label{eigenzeroper}
\end{eqnarray}

To simplify the notations , let us now 
focus on the particular state
\begin{eqnarray}
 \vert 0 > && \equiv  \vert S_1=+1,S_2=+1..,S_{L}=+1 >
\label{eigenzero}
\end{eqnarray}
whose energy
\begin{eqnarray}
E^{(0)}_{0} =- \sum_{i=1}^{L} h_i 
\label{e00}
\end{eqnarray}
 is arbitrary in the spectrum of the unperturbed Hamiltonian $H_0$,
as a consequence of the random fields of Eq. \ref{gauss}.
Let us label the other $(2^{L}-1)$ states
 by the $1 \leq k \leq {L}$ positions $(i_1,..,i_k)$
of flipped spins with respect to this reference configuration
\begin{eqnarray}
 \vert i_1,..,i_k > && \equiv  \sigma^x_{i_1} \sigma^x_{i_2} ... \sigma^x_{i_k} \vert 0 >
\label{eigenik}
\end{eqnarray}
Then the perturbed eigenstate of Eq. \ref{eigenzeroper} reads
\begin{eqnarray}
 \vert \psi^{(1)} > 
&& =  \vert 0  >
+ \sum_{k=1}^L  \sum_{1 \leq i_1< i_2..<i_k \leq {L}}
\left( \frac{ J_{i_1,i_2,..,i_k} }{ 2 \displaystyle \sum_{q=1}^k h_{i_q} } \right) \vert i_1,..,i_k >
\label{eigenzerofin}
\end{eqnarray}

 \subsection{ Reduced density matrix of the region $A$ }

To evaluate the entanglement between the two regions $A=[1,L_A]$ and 
$B=[L_A+1,L=L_A+L_B]$, it is convenient to introduce
 the following basis in each region
 with the same notations above 
\begin{eqnarray}
\vert 0>_A && \equiv  \vert S_1=+1,S_2=+1..,S_{L_A}=+1 >
\nonumber \\
 \vert i_1,..,i_k >_A && \equiv  \sigma^x_{i_1} \sigma^x_{i_2} ... \sigma^x_{i_k} \vert 0 >_A \ \ \ \ \ {\rm for } \ \ \ \ \ 1 \leq i_1< i_2..<i_k \leq {L_A}
\label{eigenA}
\end{eqnarray}
and
\begin{eqnarray}
\vert 0>_B && \equiv  \vert S_{L_A+1}=+1,S_{L+2}=+1..,S_{L}=+1 >
\nonumber \\
 \vert j_1,..,j_k >_B && \equiv  \sigma^x_{j_1} \sigma^x_{j_2} ... \sigma^x_{j_k} \vert 0 >_B \ \ \ \ \ {\rm for } \ \ \ \ \ L_A+1 \leq j_1< j_2..<j_k \leq L_A+L_B
\label{eigenB}
\end{eqnarray}

Then the eigenstate of Eq. \ref{eigenzerofin}
can be decomposed into
\begin{eqnarray}
&& \vert \psi^{(1)} > 
 = \vert 0 >_A \otimes \vert 0 >_B
\nonumber \\
&& +  \sum_{k=1}^{L_A}  \sum_{1 \leq i_1< i_2..<i_k \leq {L_A}}
\frac{ J_{i_1,i_2,..,i_k} }{ 2 \sum_{q=1}^k h_{i_q} }\vert i_1,..,i_k >_A \otimes \vert 0 >_B
\nonumber \\
&&  +  \sum_{k=1}^{L_B}  \sum_{L_A+1 \leq j_1< j_2..<k_k \leq {L_A+L_B}}
\frac{ J_{i_1,i_2,..,i_k} }{ 2 \sum_{q=1}^k h_{i_q} }
\vert 0 >_A \otimes \vert i_1,..,i_k >_B 
\nonumber \\
&& +  \sum_{k_A=1}^{L_A}  \sum_{1 \leq i_1< i_2..<i_{k_A} \leq {L_A}}
\sum_{k_B=1}^{L_B} \sum_{L_A+1 \leq j_1< j_2..<j_{k_B} \leq {L_A+L_B}}
\frac{ J_{i_1,..,i_{k_A},j_1,...j_{k_B}} }{ 2 \sum_{q=1}^{k_A} h_{i_q}
+2 \sum_{p=1}^{k_B} h_{j_p} }\vert i_1,..,i_{k_A} >_A \otimes \vert j_1,..,j_{k_B} >_B 
\label{eigenref}
\end{eqnarray}

It is thus convenient
 to use instead the following perturbed basis for the region $A$
\begin{eqnarray}
\vert \phi^A_0> && \equiv  \vert 0 >_A
+  \sum_{k=1}^{L_A}  \sum_{1 \leq i_1< i_2..<i_k \leq {L_A}}
\frac{ J_{i_1,i_2,..,i_k} }{ 2 \sum_{q=1}^k h_{i_q} }\vert i_1,..,i_k >_A 
\nonumber \\
 \vert \phi^A_{i_1,..,i_k} > && \equiv  \vert i_1,..,i_k >_A-
\frac{ J_{i_1,i_2,..,i_k} }{ 2 \sum_{q=1}^k h_{i_q} } \vert 0 >_A  
 \ \ \ \ \ {\rm for } \ \ \ \ \ 1 \leq i_1< i_2..<i_k \leq {L_A}
\label{eigenAV}
\end{eqnarray}
and the similar basis for the region $B$
\begin{eqnarray}
\vert \phi^B_0> && \equiv  \vert 0 >_B
+  \sum_{k=1}^{L_B}  \sum_{L_A+1 \leq j_1< j_2..<j_k \leq {L_A+L_B}}
\frac{ J_{j_1,j_2,..,j_k} }{ 2 \sum_{q=1}^k h_{j_q} }\vert j_1,..,j_k >_B 
\nonumber \\
 \vert \phi^B_{j_1,..,j_k} > && \equiv   \vert j_1,..,j_k >_B-
\frac{ J_{j_1,j_2,..,j_k} }{ 2 \sum_{q=1}^k h_{j_q} } \vert 0 >_B 
 \ \ \ \ \ {\rm for } \ \ \ \ \ L_A+1 \leq j_1< j_2..<j_k \leq {L_A+L_B}
\label{eigenBV}
\end{eqnarray}

Then Eq. \ref{eigenref} can be rewritten at first order in the perturbation as
\begin{eqnarray}
&& \vert \psi^{(1)} > 
 = \vert \phi^A_0 > \otimes \vert \phi^B_0 >
\nonumber \\
&& +  \sum_{k_A=1}^{L_A}  \sum_{1 \leq i_1< i_2..<i_{k_A} \leq {L_A}}
\sum_{k_B=1}^{L_B} \sum_{L_A+1 \leq j_1< j_2..<j_{k_B} \leq {L_A+L_B}}
\frac{ J_{i_1,..,i_{k_A},j_1,...j_{k_B}} }{ 2 \sum_{q=1}^{k_A} h_{i_q}
+2 \sum_{p=1}^{k_B} h_{j_p} }\vert \phi^A_{i_1,..,i_{k_A}} >
 \otimes \vert \phi^B_{j_1,..,j_{k_B}} > 
\label{eigenrefp}
\end{eqnarray}

The corresponding density matrix
\begin{eqnarray}
\rho && \equiv \frac{ \vert \psi^{(1)} > <\psi^{(1)}\vert }{<\psi^{(1)}\vert \psi^{(1)} >}
= \frac{1}{ 1+\Sigma_1 }
  ( \vert \phi^A_0 > \otimes \vert \phi^B_0 >
\nonumber \\
&& +  \sum_{k_A=1}^{L_A}  \sum_{1 \leq i_1< i_2..<i_{k_A} \leq {L_A}}
\sum_{k_B=1}^{L_B} \sum_{L_A+1 \leq j_1< j_2..<j_{k_B} \leq {L_A+L_B}}
\frac{ J_{i_1,..,i_{k_A},j_1,...j_{k_B}} }{ 2 \sum_{q=1}^{k_A} h_{i_q}
+2 \sum_{p=1}^{k_B} h_{j_p} }\vert \phi^A_{i_1,..,i_{k_A}} > \otimes \vert \phi^B_{j_1,..,j_{k_B}} > 
)
\nonumber \\ 
&& ( 
< \phi^A_0 \vert \otimes < \phi^B_0 \vert
\nonumber \\
&& +  \sum_{k_A'=1}^{L_A}  \sum_{1 \leq i_1'< i_2'..<i_{k_A'}' \leq {L_A}}
\sum_{k_B'=1}^{L_B} \sum_{L_A+1 \leq j_1'< j_2'..<j_{k_B'}' \leq {L_A+L_B}}
\frac{ J_{i_1',..,i_{k_A}',j_1',...j_{k_B'}'} }{ 2 \sum_{q=1}^{k_A'} h_{i_q'}
+2 \sum_{p=1}^{k_B'} h_{j_p'} } < \phi^A_{i_1',..,i_{k_A'}'} \vert
 \otimes < \phi^B_{j_1',..,j_{k_B'}'} \vert 
)
\label{rhoab}
\end{eqnarray}
with
\begin{eqnarray}
\Sigma_1 && \equiv \sum_{k_A=1}^{L_A}  \sum_{1 \leq i_1< i_2..<i_{k_A} \leq {L_A}}
 \sum_{k_B=1}^{L_B}  \sum_{L_A+1 \leq j_1< j_2..<j_{k_B} \leq {L_A+L_B} }
\left(\frac{ J_{i_1,..,i_{k_A},j_1,...j_{k_B}} }{ 2 \sum_{q=1}^{k_A} h_{i_q}+2 \sum_{p=1}^{k_B} h_{j_p}} \right)^2
\label{sigma1}
\end{eqnarray}
leads to the reduced density matrix for the region $A$
\begin{eqnarray}
\rho_{A} && \equiv Tr_B(\rho) = 
< \phi^B_0 \vert \rho \vert \phi^B_0 >
+ \sum_{k_B=1}^{L_B} \sum_{L_A+1 \leq j_1< j_2..<j_{k_B} \leq {L_A+L_B}}
 < \phi^B_{j_1,..,j_{k_B}} \vert \rho \vert \phi^B_{j_1,..,j_{k_B}} >
\nonumber \\
&& = \frac{1}{ 1+\Sigma_1}\vert \phi^A_0 > <\phi^A_0 \vert
\nonumber \\
&& + 
\sum_{k_A=1}^{L_A}  \sum_{1 \leq i_1< i_2..<i_{k_A} \leq {L_A}}
 \sum_{k_A'=1}^{L_A}  \sum_{1 \leq i_1'< i_2'..<i_{k_A}' \leq {L_A}}
 \frac{ R (i_1,..,i_{k_A} ; i_1',..,i_{k_A'}' ) }{ 1+\Sigma_1}
\vert \phi^A_{i_1,..,i_{k_A}} >   < \phi^A_{i_1',..,i_{k_A'}'} \vert
\label{rhoa}
\end{eqnarray}
with
\begin{eqnarray}
R (i_1,..,i_{k_A} ; i_1',..,i_{k_A'}' ) && \equiv  \sum_{k_B=1}^{L_B}
 \sum_{L_A+1 \leq j_1< j_2..<j_{k_B} \leq {L_A+L_B}}
\left(\frac{ J_{i_1,..,i_{k_A},j_1,...j_{k_B}} }{ 2 \sum_{q=1}^{k_A} h_{i_q}+2 \sum_{p=1}^{k_B} h_{j_p}} \right)
\left(\frac{ J_{i_1',..,i_{k_A}',j_1,...j_{k_B}} }{ 2 \sum_{q=1}^{k_A'} h_{i_q'}
+2 \sum_{p=1}^{k_B} h_{j_p} } \right)
\label{Raaprime}
\end{eqnarray}

In particular, the diagonal elements involve the positive coefficients
\begin{eqnarray}
D_{i_1,..,i_{k_A}} \equiv R (i_1,..,i_{k_A} ; i_1,..,i_{k_A} ) && 
\equiv  \sum_{k_B=1}^{L_B}
 \sum_{L_A+1 \leq j_1< j_2..<j_{k_B} \leq {L_A+L_B}}
\left(\frac{ J_{i_1,..,i_{k_A},j_1,...j_{k_B}} }{ 2 \sum_{q=1}^{k_A} h_{i_q}+2 \sum_{p=1}^{k_B} h_{j_p}} \right)^2
\label{diag}
\end{eqnarray}
and their sum corresponds to Eq. \ref{sigma1}
\begin{eqnarray}
\Sigma_1 && \equiv \sum_{k_A=1}^{L_A}  \sum_{1 \leq i_1< i_2..<i_{k_A} \leq {L_A}}
D_{i_1,..,i_{k_A}}
\label{sigma1d}
\end{eqnarray}

\subsection{ Entanglement spectrum }

As we will see in the next section,
the diagonal coefficients $D_{i_1,..,i_{k_A}} $ of Eq. \ref{diag}
and their sum $\Sigma_1$ of Eq. \ref{sigma1d}
are very broadly distributed with a L\'evy law of index $\mu=1/2$.
The physical meaning is that they are dominated by the few biggest terms corresponding to the smallest denominators $1/\left(2 \sum_{q=1}^{k_A} h_{i_q}+2 \sum_{p=1}^{k_B} h_{j_p} \right)^2 $ 
that can be interpreted as very rare resonances involving spins of
both regions $A$ and $B$.
The technical consequence is that 
the diagonal coefficients $D_{i_1,..,i_{k_A}} $ of Eq. \ref{diag}
and their sum $\Sigma_1$ give contributions of {\it first order } 
$O(\vert J \vert)$ in the couplings.
On the contrary, the off-diagonal coefficients of Eq. \ref{Raaprime}
involving two different denominators have a finite averaged value
 of second order $O(J^2)$ (see Appendix \ref{app_off}).

In conclusion, at first order in the couplings, the off-diagonal coefficients of Eq. \ref{Raaprime} do not contribute, so that the diagonal elements 
directly represent the entanglement spectrum, 
i.e. the $2^{L_A}$ eigenvalues of the reduced density matrix read
\begin{eqnarray}
 p_0 && =  \frac{1}{ 1+\Sigma_1}
\nonumber \\
p_{i_1,..,i_{k_A}} && =  \frac{ D_{i_1,..,i_{k_A}} }{ 1+\Sigma_1}
\ \ \ \ \  {\rm for } \ \ 1 \leq i_1< i_2..<i_{k_A} \leq {L_A}
\label{p0n}
\end{eqnarray}

To characterize the statistical properties of these weights,
it is convenient to introduce  
\begin{eqnarray}
Y_q(L_A) =  Tr_A (\rho_{A}^q  ) =  p_0^q +\sum_{k_A=1}^{L_A}  \sum_{1 \leq i_1< i_2..<i_{k_A} \leq {L_A}}p_{i_1,..,i_{k_A}}^q
= \frac{ 1 +\Sigma_q  }{ (1+\Sigma_1)^q}
\label{ipr}
\end{eqnarray}
where the sum
\begin{eqnarray}
\Sigma_q && \equiv \sum_{k_A=1}^{L_A}  \sum_{1 \leq i_1< i_2..<i_{k_A} \leq {L_A}}
D^q_{i_1,..,i_{k_A}}
\label{sigmaq}
\end{eqnarray}
generalizes Eq. \ref{sigma1d}.

Then the R\'enyi entanglement entropy of index $q$
 \begin{eqnarray}
S_q(L_A) \equiv  \frac{ \ln Y_q(L_A) }{1-q} =\frac{\ln( 1 +\Sigma_q) -q \ln (1+\Sigma_1)}{1-q}
\label{renyi}
\end{eqnarray}
allows to recover the usual entanglement entropy in the limit $q \to 1$
 \begin{eqnarray}
S_1(L_A) \equiv - Tr_A(\rho_{A} \ln \rho_{A}) =
 -p_0 \ln p_0 -\sum_{k_A=1}^{L}  \sum_{1 \leq i_1< i_2..<i_{k_A} \leq {L}} p_{i_1,..,i_{k_A}} \ln p_{i_1,..,i_{k_A}}
\label{ent}
\end{eqnarray}

In particular, to obtain the disorder-averaged value of the Renyi
entropy of Eq. \ref{renyi}
\begin{eqnarray}
\overline{ S_q } 
= \frac{\overline{\ln( 1 +\Sigma_q)} -q \overline{ \ln (1+\Sigma_1)} }{1-q}
\label{entqav}
\end{eqnarray}
one only needs to study separately the probability distributions 
of $\Sigma_q$ and $\Sigma_1$, as done in the next section.
Other statistical properties involving correlations
between $\Sigma_q$ and $\Sigma_1$
 can also be computed (see for instance Appendix \ref{app_yqav}
for the disorder-averaged values $\overline{Y_q}$ and for the variance 
of the entanglement entropy $S_1$).

\section{ Statistical properties of the entanglement spectrum }

\label{sec_ent}

In this section, we analyze the statistical properties of the entanglement spectrum obtained in Eq. \ref{p0n}.

\subsection { Probability distribution of the variable $D_{i_1,..,i_{k_A}}$ }

Let us first consider the probability distribution
$ P_{i_1,..,i_{k_A}}(D_{i_1,..,i_{k_A}})$
of the positive random variable $ D_{i_1,..,i_{k_A}}$ of Eq. \ref{diag}
by evaluating its Laplace transform at lowest order in the couplings
\begin{eqnarray}
 \overline{ e^{ - t  D_{i_1,..,i_{k_A}}  }  } && = \int_0^{+\infty} dD_{i_1,..,i_{k_A}} 
P_{i_1,..,i_{k_A}}(D_{i_1,..,i_{k_A}}) e^{-t D_{i_1,..,i_{k_A}}  }
\nonumber \\
&& =\overline{ e^{ - t  \sum_{k_B=1}^{L_B}
 \sum_{L_A+1 \leq j_1< j_2..<j_{k_B} \leq {L_A+L_B}}
\left(\frac{ J_{i_1,..,i_{k_A},j_1,...j_{k_B}} }{ 2 \sum_{q=1}^{k_A} h_{i_q}+2 \sum_{p=1}^{k_B} h_{j_p}} \right)^2 }  }
\nonumber \\
&& =\overline{ 
\prod_{k_B=1}^{L_B}
 \prod_{L_A+1 \leq j_1< j_2..<j_{k_B} \leq {L_A+L_B}}
\left[1- \left(1-e^{- t\left(\frac{ J_{i_1,..,i_{k_A},j_1,...j_{k_B}} }{ 2 \sum_{q=1}^{k_A} h_{i_q}+2 \sum_{p=1}^{k_B} h_{j_p}} \right)^2  } \right)  \right] }
\nonumber \\
&& = 
1 -  \sum_{k_B=1}^{L_B}
 \sum_{L_A+1 \leq j_1< j_2..<j_{k_B} \leq {L_A+L_B}}
\overline{ \left(1-e^{- t \left(\frac{ J_{i_1,..,i_{k_A},j_1,...j_{k_B}} }{ 2 \sum_{q=1}^{k_A} h_{i_q}+2 \sum_{p=1}^{k_B} h_{j_p}} \right)^2 } \right) }  +o(J)
\label{laplace}
\end{eqnarray}
The variable 
\begin{eqnarray}
E=2 \sum_{q=1}^{k_A} h_{i_q}+2 \sum_{p=1}^{k_B} h_{j_p}
\label{esum}
\end{eqnarray}
is a sum of $(k_A+k_B)$ Gaussian variables (Eq. \ref{gauss})
and is thus distributed with the Gaussian of zero mean and
of variance $(4 W^2 (k_A+k_B))$
\begin{eqnarray}
G_{4 W^2 (k_A+k_B)}(E) = \frac{1}{2 \sqrt { 2 (k_A+k_B)  \pi W^2 }}
 e^{- \frac{ E^2}{ 8 (k_A+k_B)  W^2} }
\label{gaussdelta}
\end{eqnarray}
So using the change of variable 
$x=t_{i_1,..,i_{k_A}} \frac{ J_{i_1,,..j_{k_B}}^2 }{E^2 } $, one obtains
\begin{eqnarray}
&& \overline{1-e^{- t \left(\frac{ J_{i_1,..,i_{k_A},j_1,...j_{k_B}} }{ 2 \sum_{q=1}^{k_A} h_{i_q}+2 \sum_{p=1}^{k_B} h_{j_p}} \right)^2 }  } 
\nonumber \\ && = \overline{ \int_{0}^{+\infty}  \frac{ dE }{\sqrt { 2 (k_A+k_B)  \pi W^2 }} e^{- \frac{ E^2}{ 8 (k_A+k_B)  W^2} } \left(1- 
e^{- t \left(\frac{ J_{i_1,..,i_{k_A},j_1,...j_{k_B}} }{ E } \right)^2  } \right) }
\nonumber \\
&& =\overline{  \vert J_{i_1,..,i_{k_A},j_1,...j_{k_B}} \vert \ \ t^{\frac{1}{2}} \frac{1}{2\sqrt { 2  (k_A+k_B) \pi W^2 } }
 \int_{0}^{+\infty}  \frac{ dx  }{  x^{\frac{3}{2}} }
 e^{- \frac{ J_{i_1,i_2,..,i_k}^2 t_A }{ 8 (k_A+k_B)  W^2 x } }\left(1- e^{- x } \right) }
\nonumber \\
&& = \overline{ \vert J_{i_1,..,i_{k_A},j_1,...j_{k_B}} \vert } \ \ t^{\frac{1}{2}} \frac{1}{2\sqrt { 2  (k_A+k_B) \pi W^2 } }
 \int_{0}^{+\infty}  \frac{ dx  }{  x^{\frac{3}{2}} } \left(1- e^{- x } \right) +o(J_{i_1,..,i_{k_A},j_1,...j_{k_B}})
\nonumber \\
&& =\overline{ \vert J_{i_1,..,i_{k_A},j_1,...j_{k_B}} \vert }\ \ t^{\frac{1}{2}} \frac{1}{\sqrt { 2 (k_A+k_B)  W^2 } }
  +o(J_{i_1,..,i_{k_A},j_1,...j_{k_B}})
\label{laplaceab}
\end{eqnarray}
so that Eq \ref{laplace}  becomes at first order in the couplings $J_{i_1,..,i_{k_A},j_1,...j_{k_B}} $
\begin{eqnarray}
&& \overline{ e^{ - t_{i_1,..,i_{k_A}} D_{i_1,..,i_{k_A}}  }  }
\nonumber \\
&& = 
1 -  t^{\frac{1}{2}}  \sum_{k_B=1}^{L_B}
 \sum_{L_A+1 \leq j_1< j_2..<j_{k_B} \leq {L_A+L_B}}
\overline{  \vert J_{i_1,..,i_{k_A},j_1,...j_{k_B}} \vert } \ \ \frac{1}{\sqrt { 2 (k_A+k_B)  W^2 } }  +o(J)
\nonumber \\
&& = e^{  
-   t^{\frac{1}{2}}
 \sum_{k_B=1}^{L_B}
 \sum_{L_A+1 \leq j_1< j_2..<j_{k_B} \leq {L_A+L_B}}
\overline{ \vert J_{i_1,..,i_{k_A},j_1,...j_{k_B}} \vert } \ \frac{1}{\sqrt { 2 (k_A+k_B)  W^2 } }  +o(J) }
\label{laplaceres}
\end{eqnarray}

It is thus convenient to introduce the notation
\begin{eqnarray}
\Omega_{i_1,..,i_{k_A}}  \equiv \sum_{k_B=1}^{L_B}
 \sum_{L_A+1 \leq j_1< j_2..<j_{k_B} \leq {L_A+L_B}}
\overline{ \vert J_{i_1,..,i_{k_A},j_1,...j_{k_B}} \vert } \ \frac{1}{\sqrt { 2 (k_A+k_B)  W^2 } }
\label{omega}
\end{eqnarray}
and the L\'evy positive stable-law of index $\mu=1/2$ and of parameter $\Omega$ 
\begin{eqnarray}
L_{\frac{1}{2};\Omega}( D) && \equiv 
 \frac{\Omega}{2 \sqrt{\pi} D^{\frac{3}{2}} } e^{- \frac{\Omega^2}{4 D} }
\label{levy}
\end{eqnarray}
with its Laplace transform
\begin{eqnarray}
{\hat L}_{\frac{1}{2},\Omega}( t ) && \equiv
 \int_0^{+\infty} dD L_{\frac{1}{2};\Omega}( D) e^{- t D}  = 
 e^{- \Omega t^{\frac{1}{2} } }
\label{levylaplace}
\end{eqnarray}
The parameter $\Omega$ directly measures the weight of the singularity
 in $t^{\frac{1}{2}} $ of the Laplace transform,
and thus also the weight of the power-law behavior at large $D$ of the distribution
of Eq. \ref{levy}. L\'evy variables \cite{levy} actually appear very often in 
the studies of disordered systems (see for instance \cite{jpbreview,Der,Der_Fly,us_critiweights} for more details on their properties).

Then Eq. \ref{laplaceres} means that at lowest order in the couplings $J$,
the variable $ D_{i_1,..,i_{k_A}} $ is distributed with 
the  L\'evy law $L_{\frac{1}{2};\Omega_{i_1,..,i_{k_A}}}$
\begin{eqnarray}
P_{i_1,..,i_{k_A}}(D_{i_1,..,i_{k_A}}) = L_{\frac{1}{2};\Omega_{i_1,..,i_{k_A}}}( D_{i_1,..,i_{k_A}}) 
\label{levyomega}
\end{eqnarray}

\subsection{ Probability distribution of $\Sigma_1$}

The probability distribution
$ P_{1}(\Sigma_1)$
of the positive random variable $\Sigma1$ of Eq. \ref{sigma1d}
can be evaluated similarly via its Laplace transform
\begin{eqnarray}
 \overline{ e^{ - t \Sigma_1  }  } && \equiv \int_0^{+\infty} d\Sigma_1 
P_1(\Sigma_1) e^{-t \Sigma_1  }
\nonumber \\
&& =\overline{ e^{ - t \sum_{k_A=1}^{L_A}  \sum_{1 \leq i_1< i_2..<i_{k_A} \leq {L_A}} \sum_{k_B=1}^{L_B}
 \sum_{L_A+1 \leq j_1< j_2..<j_{k_B} \leq {L_A+L_B}}
\left(\frac{ J_{i_1,..,i_{k_A},j_1,...j_{k_B}} }{ 2 \sum_{q=1}^{k_A} h_{i_q}+2 \sum_{p=1}^{k_B} h_{j_p}} \right)^2 }  }
\nonumber \\
&& =\overline{ \prod_{k_A=1}^{L_A}  \prod_{1 \leq i_1< i_2..<i_{k_A} \leq {L_A}}
\prod_{k_B=1}^{L_B}
 \prod_{L_A+1 \leq j_1< j_2..<j_{k_B} \leq {L_A+L_B}}
\left[1- \left(1-e^{- t\left(\frac{ J_{i_1,..,i_{k_A},j_1,...j_{k_B}} }{ 2 \sum_{q=1}^{k_A} h_{i_q}+2 \sum_{p=1}^{k_B} h_{j_p}} \right)^2  } \right)  \right] }
\nonumber \\
&& = 
1 - \sum_{k_A=1}^{L_A}  \sum_{1 \leq i_1< i_2..<i_{k_A} \leq {L_A}} \sum_{k_B=1}^{L_B}
 \sum_{L_A+1 \leq j_1< j_2..<j_{k_B} \leq {L_A+L_B}}
\overline{ \left(1-e^{- t \left(\frac{ J_{i_1,..,i_{k_A},j_1,...j_{k_B}} }{ 2 \sum_{q=1}^{k_A} h_{i_q}+2 \sum_{p=1}^{k_B} h_{j_p}} \right)^2 } \right) }  +o(J)
\label{laplacesigma1}
\end{eqnarray}

Eq. \ref{laplaceab} then yields that
\begin{eqnarray}
\overline{ e^{-t \Sigma_1} } && =
1 - t^{\frac{1}{2}} \sum_{k_A=1}^{L_A}  \sum_{1 \leq i_1< i_2..<i_{k_A} \leq {L_A}} \sum_{k_B=1}^{L_B}
 \sum_{L_A+1 \leq j_1< j_2..<j_{k_B} \leq {L_A+L_B}}
\overline{ \vert J_{i_1,..,i_{k_A},j_1,...j_{k_B}} \vert }\ \ \frac{1}{\sqrt { 2 (k_A+k_B)  W^2 } }  +o(J)
\nonumber \\
&& =  e^{  - \Omega_1  t^{\frac{1}{2}} +o(J) }   
\label{lapsigma1d}
\end{eqnarray}
with
\begin{eqnarray}
\Omega_1 && \equiv \sum_{k_A=1}^{L_A}  \sum_{1 \leq i_1< i_2..<i_{k_A} \leq {L_A}}
 \Omega_{i_1,..,i_{k_A}}  
\nonumber \\ &&
= \sum_{k_A=1}^{L_A}  \sum_{1 \leq i_1< i_2..<i_{k_A} \leq {L_A}}
\sum_{k_B=1}^{L_B}
 \sum_{L_A+1 \leq j_1< j_2..<j_{k_B} \leq {L_A+L_B}}
\vert J_{i_1,..,i_{k_A},j_1,...j_{k_B}} \vert \ \frac{1}{\sqrt { 2 (k_A+k_B)  W^2 } }
\label{omega1}
\end{eqnarray}
Eq. \ref{lapsigma1d} means that at lowest order in the couplings $J$,
the variable $\Sigma_1$ is distributed with 
the  L\'evy law $L_{\frac{1}{2};\Omega_1}$ (Eqs \ref{levy} and \ref{levylaplace})
of index $\mu=1/2$ and of parameter $\Omega_1$
\begin{eqnarray}
P_{1}(\Sigma_1) = L_{\frac{1}{2};\Omega_1}(\Sigma_1 ) 
\label{levysigma1}
\end{eqnarray}

In particular, using the identity
\begin{eqnarray}
\ln (1+ \Sigma_1) = \int_0^{+\infty} \frac{dt}{t} e^{-t} \left( 1-e^{- t \Sigma_1} \right)
\label{logrepresentation}
\end{eqnarray}
one obtains
\begin{eqnarray}
\overline{ \ln (1+\Sigma_1)} && = \int_0^{+\infty} d\Sigma_1 P_{1}(\Sigma_1)  \ln (1+\Sigma_1) \nonumber \\ &&
=  \int_0^{+\infty} \frac{dt}{t} e^{-t} \left( 1- \overline{ e^{- t \Sigma_1} } \right)
 \nonumber \\ &&
=  \int_0^{+\infty} \frac{dt}{t} e^{-t} \Omega_1 t^{\frac{1}{2}}  +o(J)
 \nonumber \\ &&
= \Omega_1 \sqrt{\pi} +o(J)
\label{lnsigma1}
\end{eqnarray}

\subsection{ Probability distribution of the weight $p_0$ } 

The probability distribution ${\cal P}(p_0)$ 
of the weight $0 \leq p_0 \leq 1$ of the unperturbed state (Eq \ref{p0n})
\begin{eqnarray}
 p_0 && =  \frac{1}{ 1+\Sigma_1}
\label{p0unper}
\end{eqnarray}
can be obtained directly from Eq \ref{levysigma1}
by the change of variable
\begin{eqnarray}
{\cal P}(p_0) && = \frac{1}{p_0^2}  L_{\frac{1}{2};\Omega_1}\left(\frac{1}{p_0}-1 \right) 
= \frac{\Omega_1 p_0^{\frac{1}{2}}}{2 \sqrt{\pi} (1-p_0)^{\frac{3}{2}} }
 e^{- \frac{\Omega_1^2 p_0}{4 (1-p_0)} }
\label{probap0}
\end{eqnarray}
In particular, its averaged value 
\begin{eqnarray}
\int_0^1 d p_0 p_0 {\cal P}(p_0) && = 1 - \Omega_1  
e^{ \frac{\Omega_1^2 }{4 } }
\int_{\frac{\Omega_1}{2}}^{+\infty} dt e^{-t^2}
\label{avp0}
\end{eqnarray}
is close to unity for small $\Omega_1$
\begin{eqnarray}
\int_0^1 d p_0 p_0 {\cal P}(p_0) && = 1 - \frac{\sqrt{\pi}}{2} \Omega_1+O(\Omega_1^2)  
\label{avp0dv1}
\end{eqnarray}

\subsection{ Probability distribution of $\Sigma_q$} 

Let us now consider the variable $\Sigma_q$ of Eq. \ref{sigmaq}.

Here one has to distinguish two regions for the index $q$ :

(i) For $0<q<\frac{1}{2} $, the moment of order $q$ of $ D_{i_1,..,i_{k_A}}$
distributed with the L\'evy law of Eq. \ref{levy}
is finite and reads
\begin{eqnarray}
\overline{D^q_{i_1,..,i_{k_A}} } = \int_0^{+\infty}  
dD\frac{\Omega_{i_1,..,i_{k_A}}}{2 \sqrt{\pi} }D^{q-\frac{3}{2}}  
e^{- \frac{\Omega_{i_1,..,i_{k_A}}^2}{4 D} } =
 \frac{\Omega_{i_1,..,i_{k_A}}^{2q} \Gamma \left(
\frac{1}{2}-q  \right) }{4^q \sqrt{\pi}}
\label{levyq}
\end{eqnarray}
As a consequence, the average of $\Sigma_q $ is also finite and given by
\begin{eqnarray}
\overline{\Sigma_q  } =
\sum_{k_A=1}^{L} 
 \sum_{1 \leq i_1< i_2..<i_{k_A} \leq {L}}\overline{D^q_{i_1,..,i_{k_A}} } 
 = \frac{ \Gamma \left(
\frac{1}{2}-q  \right) }{4^q \sqrt{\pi}}
\sum_{k_A=1}^{L} 
 \sum_{1 \leq i_1< i_2..<i_{k_A} \leq {L}} \Omega_{i_1,..,i_{k_A}}^{2q}
\label{avsigmaq}
\end{eqnarray}
so that at lowest order in the couplings one has
\begin{eqnarray}
\overline{ \ln (1+\Sigma_q)} && \simeq \overline{\Sigma_q  } 
= \frac{ \Gamma \left(
\frac{1}{2}-q  \right) }{4^q \sqrt{\pi}}
\sum_{k_A=1}^{L} 
 \sum_{1 \leq i_1< i_2..<i_{k_A} \leq {L}} \Omega_{i_1,..,i_{k_A}}^{2q}
 +o(\vert J \vert^{2q})
\label{lnsigmaqsmall}
\end{eqnarray}

(ii) For $q>\frac{1}{2} $, the average of $\Sigma_q $ is infinite,
and one needs to estimate the Laplace transform
of its probability distribution $P_q(\Sigma_q)$ as above
\begin{eqnarray}
\overline{ e^{- t \Sigma_q } } &&\equiv \int_0^{+\infty} d\Sigma_q 
P_q(\Sigma_q) e^{-t \Sigma_q  } =\prod_{k_A=1}^{L} 
 \prod_{1 \leq i_1< i_2..<i_{k_A} \leq {L}}
\left[ \overline{ e^{- t  D^q_{i_1,..,i_{k_A}} } } \right]
\nonumber \\ &&
= \prod_{k_A=1}^{L} 
 \prod_{1 \leq i_1< i_2..<i_{k_A} \leq {L}}
 \overline{ 1- \left( 1- e^{- t  D^q_{i_1,..,i_{k_A}} } \right) }
\nonumber \\ &&
= 1- \sum_{k_A=1}^{L} 
 \sum_{1 \leq i_1< i_2..<i_{k_A} \leq {L}} \overline{ \left( 1- e^{- t  D^q_{i_1,..,i_{k_A}} } \right) } +o(J)
\label{lapsigmaq}
\end{eqnarray}

Using Eq. \ref{levy} and the change of variable $x= t  D^q  $, one obtains
\begin{eqnarray}
\overline{ \left( 1- e^{- t  D^q } \right) } && =
\int_0^{+\infty}  
dD \frac{\Omega}{2 \sqrt{\pi} D^{\frac{3}{2}} } e^{- \frac{\Omega^2}{4 D} }
\left( 1- e^{- t  D^q } \right)
\nonumber \\ &&
=\frac{\Omega}{2 q \sqrt{\pi} } t^{\frac{1}{2q}} 
\int_0^{+\infty}    \frac{dx}{x^{1+\frac{1}{2q}} } (1-e^{-x}) +o(J)
\nonumber \\ &&
=\frac{\Omega \sqrt{\pi} }{  \sin \left(\frac{\pi}{2q} \right) \Gamma \left(\frac{1}{2q} \right)} t^{\frac{1}{2q}} 
 +o(J)
\label{levyqlap}
\end{eqnarray}
so that Eq. \ref{lapsigmaq} becomes
\begin{eqnarray}
\overline{ e^{- t \Sigma_q } } && 
= 1-\frac{ \sqrt{\pi} }{  \sin \left(\frac{\pi}{2q} \right) \Gamma \left(\frac{1}{2q} \right)} t^{\frac{1}{2q}}  \sum_{k_A=1}^{L} 
 \sum_{1 \leq i_1< i_2..<i_{k_A} \leq {L}} \Omega_{i_1,..,i_{k_A}}
 +o(J)
\nonumber \\ &&
= e^{-\frac{ \sqrt{\pi} \Omega_1 }{  \sin \left(\frac{\pi}{2q} \right) \Gamma \left(\frac{1}{2q} \right)} t^{\frac{1}{2q}}  }  +o(J)
\label{lapsigmaqres}
\end{eqnarray}
It is thus convenient to introduce the notation
\begin{eqnarray}
\Omega_q \equiv \frac{ \sqrt{\pi} \Omega_1 }{  \sin \left(\frac{\pi}{2q} \right) \Gamma \left(\frac{1}{2q} \right)}
\label{omegaq}
\end{eqnarray}
that generalizes Eq. \ref{omega1}, and 
 the L\'evy positive stable-law $L_{\mu;\Omega}$ of index $0<\mu<1$ and of parameter $\Omega$ defined by its Laplace transform
\begin{eqnarray}
{\hat L}_{\mu,\Omega}( t ) && \equiv
 \int_0^{+\infty} dD L_{\mu;\Omega}( D) e^{- t D}  = 
 e^{- \Omega t^{\mu} }
\label{levylaplacemu}
\end{eqnarray}

Eq. \ref{lapsigmaqres}
 means that for $q>\frac{1}{2} $,
the variable $\Sigma_q $ is distributed with the L\'evy stable law 
$L_{\mu_q;\Omega_q}$ of index $\mu_q=\frac{1}{2q}$ and parameter $\Omega_q$
\begin{eqnarray}
P_{q}(\Sigma_q) = L_{\frac{1}{2q};\Omega_q}(\Sigma_q ) 
\label{levysigmaq}
\end{eqnarray}
that generalizes Eq \ref{levysigma1}.
In particular, using the identity of Eq. \ref{lnsigma1} for $\Sigma_q$,
 one obtains
\begin{eqnarray}
\overline{ \ln (1+\Sigma_q)} && = \int_0^{+\infty} d\Sigma_q P_{q}(\Sigma_q)  \ln (1+\Sigma_q) \nonumber \\ &&
=  \int_0^{+\infty} \frac{dt}{t} e^{-t} \left( 1- \overline{ e^{- t \Sigma_q} } \right)
 \nonumber \\ &&
=  \int_0^{+\infty} \frac{dt}{t} e^{-t} \Omega_q t^{\frac{1}{2q}}  +o(J)
 \nonumber \\ &&
= \Omega_q \Gamma \left( \frac{1}{2q} \right) +o(J)
 \nonumber \\ &&
= \Omega_1 \frac{ \sqrt{\pi} }{  \sin \left(\frac{\pi}{2q} \right)}  +o(J)
\label{lnsigmaq}
\end{eqnarray}

\subsection{ Disorder-averaged values of the Renyi entropies }

The disorder-averaged entanglement entropy of Eq. \ref{entqav}
can be obtained from the previous results :

(i) in the region $0<q<1/2$, 
 the contribution of Eq. \ref{lnsigmaqsmall} of order $\vert J \vert^{2q}$
dominates over the contribution of Eq. \ref{lnsigma1}, so that
the leading order reads
\begin{eqnarray}
\overline{ S_q } 
= \frac{\Gamma \left(
\frac{1}{2}-q  \right)  }{(1-q)4^q \sqrt{\pi}} \sum_{k_A=1}^{L} 
 \sum_{1 \leq i_1< i_2..<i_{k_A} \leq {L}}
\Omega_{i_1,..,i_{k_A}}^{2q}  +o(\vert J \vert^{2q})
\label{entqavqsmall}
\end{eqnarray}

(ii) in the region $q>\frac{1}{2} $ one obtains using Eq. \ref{lnsigma1}
and Eq. \ref{lnsigmaq}  
\begin{eqnarray}
\overline{ S_q } 
= \frac{\Omega_1 \sqrt{\pi}  }{1-q}\left[
 \frac{ 1 }{  \sin \left(\frac{\pi}{2q} \right)} -q \right]+o(J)
\label{entqavqlarge}
\end{eqnarray}
and in particular in the limit $q \to 1$
\begin{eqnarray}
\overline{ S_1 } 
=  \Omega_1 \sqrt{\pi} +o(J)
\label{entavq1}
\end{eqnarray}

In conclusion, the parameter $\Omega_1$ of Eq. \ref{omega1}
directly represents the scale of the disorder-averaged entanglement entropy of Eq. \ref{entavq1}.
 As shown in Eq. \ref{entvar} of Appendix \ref{app_yqav}, the scale $\Omega_1$ also governs
the variance of the entanglement entropy $ S_1 $.

\section{ Scaling of the entanglement entropy with the length $L_A$ }

\label{sec_omega}

In this section, we study how the scale $\Omega_1$ (Eq. \ref{omega1})
 of the disorder-averaged entanglement entropy of Eq. \ref{entavq1}
depends on the length $L_A$ of the region $A$.
Here we need to make more precise assumptions on the couplings $J_{i_1,..,i_k} $.

\subsection{ Statistical properties of the couplings $J_{i_1,..,i_k} $ }

As explained in Section \ref{sec_toy}, 
within the {\it toy model } of Eq. \ref{Htot} that we consider for the MBL case,
the couplings $J_{i_1,..,i_k} $ involved in the perturbation of Eq. \ref{Vperturbation} are the analog of the long-ranged hoppings of Eq. \ref{prbm} for the Anderson model. In the Anderson case, the hopping between two points depends only on the distance, but in the MBL case, we have to choose the properties for all the couplings $J_{i_1,..,i_k}  $.
To simplify the discussion, let us make the simplest possible choice and 
assume that
 the couplings $J_{i_1,..,i_k} $ are independent Gaussian random variables
\begin{eqnarray}
P(J_{i_1,..,i_k}) = \frac{1}{\sqrt { 2  \pi \Delta^2(i_k-i_1) }}
 e^{- \frac{ J_{i_1,..,i_k}^2}{2  \Delta^2(i_k-i_1)} }
\label{gaussJ}
\end{eqnarray}
of zero mean
\begin{eqnarray}
\overline{ J_{i_1,..,i_k} } =0 
\label{javzero}
\end{eqnarray}
and of variance depending only on the spatial 
range $r \equiv i_k-i_1$, i.e. on the distance between the two extremal spins
\begin{eqnarray}
\overline{ J^2_{i_1,..,i_k} } = \Delta^2(r \equiv i_k-i_1) 
\label{jvar}
\end{eqnarray}
In addition, we consider the following size dependence
 with three parameters
\begin{eqnarray}
 \Delta(r ) = \frac{v}{r^a} 2^{-b r} 
\label{jr}
\end{eqnarray}
where $b$ governs the exponential decay, $a$ governs the power-law prefactor,
and $v$ is the small amplitude of the perturbation theory described
in the previous sections.
From the exponential decay of the level spacing of Eq. \ref{levelspacing},
one may already guess that the critical value for the exponential decay
will be $b_c=1$, as found indeed below. 

Since the perturbation $V$ of Eq. \ref{Htot}
is a part of the full Hamiltonian of Eq. \ref{Vperturbation},
we wish to impose that it remains extensive with respect to the number of spins.
This means that the local field in the $x$ direction 
on a given spin, for instance the first one
\begin{eqnarray}
B^x_{1} =  - \sum_{k=2}^{L}  \sum_{1 < i_2..<i_k \leq {L}} J_{1,..,i_k} S^x_{i_2} ... S^x_{i_k}
\label{bloc}
\end{eqnarray}
should remain a finite random variable as $L \to +\infty$.
Its variance may be evaluated as
\begin{eqnarray}
\overline{ (B^x_{1})^2 } && \simeq 
\sum_{k=2}^{L}  \sum_{1 < i_2..<i_k \leq {L}} \overline{ J_{1,..,i_k}^2 } 
= \sum_{k=2}^{L}  \sum_{1 < i_2..<i_k \leq {L}} \Delta^2(i_k-1)
\label{bvar}
\end{eqnarray}
When $i_k=1+r$ is fixed, the number of ways to place 
the remaining $q=(k-2)$ points $(i_2,..,i_{k-1})$
into the $(i_k-2)=r-1$ possible positions $(i_1+1,..,i_k-1)$
is given by the binomial coefficient $\binom {r-1} {q }$,
so that Eq. \ref{bloc} yields
\begin{eqnarray}
\overline{ (B^x_{1})^2 } && \simeq 
   \sum_{r=1} \Delta^2(r)
 \sum_{q=0}^{r-1}   \binom {r-1} {q }
 =  \sum_{r=1} \Delta^2(r) 2^{r-1}
\label{bvarfin}
\end{eqnarray}

So the extensivity of the perturbation requires the convergence of the sum
\begin{eqnarray}
\sum_{r=1}^{+\infty}  \Delta^2(r) 2^{r} < + \infty
\label{extensive}
\end{eqnarray}
With the form of Eq. \ref{jr}, the convergence of the sum
\begin{eqnarray}
\sum_{r=1}^{+\infty}  \Delta^2(r) 2^{r} = v^2 \sum_{r=1}^{+\infty} 
\frac{1}{r^{2a}} 2^{(1-2 b) r}  
\label{Vextensivesg}
\end{eqnarray}
requires either (i) $b>1/2$  , or (ii)  $b=1/2$ with $a>1/2$.

\subsection{ Study of the scale $\Omega_1$ }

The average value of the absolute value of the coupling
distributed with Eq. \ref{gaussJ}
\begin{eqnarray}
\overline{ \vert J_{i_1,..,i_k} \vert } = 
\int_0^{+\infty} dJ J \frac{2 }{\sqrt { 2  \pi \Delta^2(i_k-i_1) }}
 e^{- \frac{ J^2}{2  \Delta^2(i_k-i_1)} } = \sqrt{\frac{2}{\pi}}  \Delta(i_k-i_1)
\label{gaussJav}
\end{eqnarray}
is also governed by the scale $\Delta(r)$ as a function of the spatial range $r$.

As a consequence, the scale $\Omega_1$ of Eq. \ref{omega1}
can be evaluated as above, i.e. once the extreme points $i_1$ and $j_{k_B}$
have been chosen, one has binomial coefficients
to take into account the number of ways
to place the $(k_A-1)$ points $(i_2,..,i_{k_A})$ in the
$(L_A-i_1)$ remaining possible positions $(i_1+1,..,L_A)$,
and to place the $(k_B-1)$ points $(j_1,..,j_{k_B-1})$ in the
$(j_{k_B}-L_A-1)$ remaining possible positions $(L_A+1,..,j_{k_B}-1)$
\begin{eqnarray}
 \Omega_1  && = \sum_{k_A=1}^{L_A}  \sum_{1 \leq i_1< i_2..<i_{k_A} \leq {L_A}}
\sum_{k_B=1}^{L_B}
 \sum_{L_A+1 \leq j_1< j_2..<j_{k_B} \leq {L_A+L_B}}
\overline{ \vert J_{i_1,..,i_{k_A},j_1,...j_{k_B}} \vert }
\ \frac{1}{\sqrt { 2 (k_A+k_B)  W^2 } }
\nonumber \\
&& = \frac{ \sqrt{2}}{\pi W } 
\sum_{i_1=1}^{L_A}
 \sum_{j_{k_B} =L_A+1}^{L_A+L_B}
  \Delta(j_{k_B}-i_1)
 \sum_{k_A=1}^{L_A-i_1+1}  
\sum_{k_B=1}^{ j_{k_B} -L_A}  \frac{1}{\sqrt {  (k_A+k_B)  } }
 { \binom { L_A-i_1 }{ k_A-1 }  }
 { \binom {j_{k_B}-L_A-1 }{ k_B-1 } } 
 \nonumber \\
&& = \frac{ \sqrt{2}}{\pi W } 
\sum_{n=0}^{L_A-1}
 \sum_{m=0}^{L_B-1}
  \Delta(1+n+m)
 \sum_{k_A=1}^{n+1}  
\sum_{k_B=1}^{ m+1}  \frac{1}{\sqrt {  (k_A+k_B)  } }
 { \binom { n }{ k_A-1 }  }
 { \binom { m }{ k_B-1 } } 
\label{omega1av}
\end{eqnarray}
where $n=L_A-i_1$ and $m=j_{k_B}-L_A-1$ represent the distances
with respect to the frontier between the regions $A$ and $B$.

Of course the scale $\Omega_1$ has always a finite contribution $O(1)$
coming from the finite distances $(n,m)$ of order $O(1)$
with respect to the frontier.
Now we wish to evaluate the contribution of large distances $(n,m)$.
Using the identity
\begin{eqnarray}
 \frac{1}{\sqrt{ k_A+k_B}}  = \frac{1}{\sqrt \pi}
 \int_0^{+\infty} du u^{-\frac{1}{2}} e^{-(k_A+k_B) u} 
\label{identitydemi}
\end{eqnarray}
one obtains
\begin{eqnarray}
&& \sum_{k_A=1}^{n+1}  
\sum_{k_B=1}^{ m+1}  \frac{1}{\sqrt {  (k_A+k_B)  } }
 { \binom { n }{ k_A-1 }  }
 { \binom { m }{ k_B-1 } } 
\nonumber \\ && =\frac{1}{\sqrt \pi}
 \int_0^{+\infty} du u^{-\frac{1}{2}}
\left[ \sum_{k_A=1}^{n+1}   { \binom{ n }{ k_A-1 }}   e^{-k_A u} 
\right]
\left[
\sum_{k_B=1}^{ m+1}  
 { \binom{ m }{ k_B-1 } }   e^{-k_B u} \right]
\nonumber \\
&& = \frac{1}{\sqrt \pi}
 \int_0^{+\infty} du u^{-\frac{1}{2}} e^{-2 u} \left[1+   e^{- u} \right]^{1+n+m}
\nonumber \\
&& = \frac{1}{\sqrt \pi}
 \int_0^{+\infty} du u^{-\frac{1}{2}} e^{-2 u} \left[2- (1-   e^{- u}) \right]^{1+n+m}
\nonumber \\
&& =2^{1+n+m} \frac{1}{\sqrt \pi}
 \int_0^{+\infty} du u^{-\frac{1}{2}} e^{-2 u} \left[1- \frac{(1-   e^{- u})}{2} \right]^{1+n+m}
\nonumber \\
&& \opsimeq_{n+m \to +\infty} 2^{1+n+m} \frac{\sqrt 2}{\sqrt {n+m} }
\label{omega1avsuite}
\end{eqnarray}
As could have been anticipated, this leading behavior means 
that the sum is dominated by the regions $k_A \sim \frac{n}{2}$
and $k_B  \sim \frac{m}{2}$ that maximize the binomial coefficients,
i.e. the dominant resonances on the large distance $(n+m)$ are extensive
resonances involving a number of spins of order $k_A+k_B \simeq \frac{n+m}{2}$.

As a consequence, after the multiplication
by $\Delta(1+n+m) $ with Eq. \ref{jr},
one obtains the asymptotic behavior
\begin{eqnarray}
 \Delta(1+n+m) \sum_{k_A=1}^{n+1}  
\sum_{k_B=1}^{ m+1}  \frac{1}{\sqrt {  (k_A+k_B)  } }
 { \binom { n }{ k_A-1 }  }
 { \binom { m }{ k_B-1 } } 
&&\oppropto_{n+m \to +\infty}  \frac{v}{(n+m)^{a+\frac{1}{2}}} 2^{- (1-b) (n+m)} 
\label{cold}
\end{eqnarray}
that we need to integrate over $n$ and $m$ to obtain 
the scale $\Omega_1$ of Eq. \ref{omega1av}.

So we arrive at the following conclusions :

(i) for $b>1$, there is an exponential convergence $2^{- (1-b) (n+m)}$
 at large distance, so the Many-Body-Localized phase
is stable and displays a finite entanglement entropy.

(ii) for $b_c=1$, there is no exponential factor anymore in Eq. \ref{cold},
but only the power-law $\frac{1}{(n+m)^{a+\frac{1}{2}}} $. 
We wish that the integral over $m$ converges in order to have a well-defined
thermodynamic limit $L_B \to +\infty$ for the region $B$ :
this is the case for $a>1/2$.
Then for $ 1/2<a<\frac{3}{2}$, the scale $\Omega_1$ is dominated by
the contribution of the long distance
\begin{eqnarray}
 \Omega_1 \left(b_c=1;1/2<a<\frac{3}{2}\right) 
&&\propto v \sum_{n=1}^{L_A} \frac{1}{ \left( a-\frac{1}{2} \right) n^{a-\frac{1}{2}}}
\nonumber \\
&&\propto \frac{ v }{\left( \frac{3}{2}-a \right)\left( a-\frac{1}{2} \right)}   L_A^{\frac{3}{2}-a}
\label{omega1criti}
\end{eqnarray}

\subsection{ Entanglement growth at criticality }

In summary, for the value $b_c=1$ in Eq. \ref{jr} for the coupling
that matches exactly the exponential decay of the level spacing of Eq. \ref{levelspacing}, we have obtained that, 
as the parameter $a$ of the power-law of Eq. \ref{jr}
varies in the interval $1/2<a<\frac{3}{2} $,
it is possible to construct a critical state with an entanglement growth
governed by the exponent
\begin{eqnarray}
0<\alpha=\frac{3}{2}-a < 1
\label{alpha}
\end{eqnarray}
anywhere between the area law $\alpha=0$ and the volume law $\alpha=1$.

Of course within the present approach, we have to stop at the critical point
and we have no access to the delocalized phase, but as recalled in the Introduction near Eq. \ref{fssdeloc}, the values $0<\alpha<1$ are only possible if
the transition is towards a delocalized non-ergodic phase \cite{grover}.

\subsection{ Correlation length exponent $\nu$ of the MBL phase }

In the MBL phase $b<b_c=1$, 
the exponential convergence $2^{- (1-b) (n+m)}$ in Eq. \ref{cold}
corresponds to the correlation length
\begin{eqnarray}
\xi = \frac{1}{b-1}
\label{xi}
\end{eqnarray}
that diverges near the transition $b \to b_c=1$
with the simple correlation length exponent 
\begin{eqnarray}
\nu =1
\label{nu}
\end{eqnarray}
This value simply reflects the crossing of the exponential decay of the level spacing of Eq. \ref{levelspacing} and of  exponential decay of the couplings
of Eq. \ref{jr}. It coincides with the 
exact correlation length exponent $\nu_{loc}=1$ for the Anderson localization 
on the Bethe lattice \cite{kunz,mirlin_fyodorov,us_andersontreeTW},
where the Hilbert space also grows exponentially with the distance.

Since this value $\nu=1$ does not satisfy the usual Harris \cite{harris}
or Chayes-Chayes-Fisher-Spencer \cite{chayes} inequality
\begin{eqnarray}
\nu  \geq \frac{2}{d} 
\label{harris}
\end{eqnarray}
that has been rediscussed recently for the specific case of the MBL transition
\cite{harrisMBL}, it is important to understand why.
The Harris inequality of Eq. \ref{harris} is based on the fact
that in a volume $L^d$ of a disordered sample, there are of order
$L^d$ random variables defining the disorder, so that 
there will be fluctuations of order $L^{\frac{d}{2}}$
as a consequence of the Central Limit Theorem.
In the present model for instance, 
the system of $L$ spins contains $L$ random fields $h_i$ (Eq. \ref{gauss}).
However, the MBL transition is very different from other
 types of phase transitions, because it is a transition concerning 
an individual arbitrary eigenstate in the middle of the spectrum,
in our case the state $\vert 0 >$ of Eq. \ref{eigenzero}.
Then the full enumeration of possible 
resonances involve the other $(2^L-1)$ states
of energies
\begin{eqnarray}
E^{(0)}_{S_1,..,S_{L}} =- \sum_{i=1}^{L} h_i S_i
\label{e0dicussion}
\end{eqnarray}
So in some sense, the eigenstate $\vert 0 >$
does not see only the $L$ random variables $h_i$,
but it sees effectively  the $(2^L-1)$ other random energies
that are build from the $L$ variables $h_i$ by the 
various choices of the spin values $S_i$,
as can be seen also in the exponentially small
level spacing of Eq. \ref{levelspacing}.

As a consequence, independently of the strong disorder limit
 considered in the present paper, we believe that the correlation length exponent
$\nu$ has no reason to satisfy the inequality of Eq. \ref{harris},
so that there is no theoretical inconsistency
in the numerical works where the correlation length exponent $\nu$
is found to violate the Harris inequality \cite{kjall,alet}.

\section{ Multifractal statistics of the entanglement spectrum}

\label{sec_multif}

As in Anderson transitions where the critical point
 is characterized by multifractal eigenfunctions \cite{mirlinrevue},
one expects that the MBL transition is related to some multifractal properties
 \cite{luca_mbl,santos,fradkin,serbyn,mbl_dyson,c_mbldysonbrownian,c_mblstrongmultif}. Besides the entanglement entropy described above, it is thus interesting to
characterize the statistics of the whole entanglement spectrum of Eq. \ref{p0n}, both in the Many-Body-Localized phase $b>1$ and at the critical point $b_c=1$.  
 
\subsection{ Multifractal exponents $\tau(q)$  }

In the region $0<q<1/2$, the disorder-averaged value of the Renyi entropy
of Eq. \ref{entqavqsmall} can be evaluated
by generalizing the previous calculations of Section IV B
to obtain
\begin{eqnarray}
\overline{ S_q } 
&& = \frac{\Gamma \left(
\frac{1}{2}-q  \right)  }{(1-q)4^q \sqrt{\pi}} \sum_{k_A=1}^{L} 
 \sum_{1 \leq i_1< i_2..<i_{k_A} \leq {L}}
\Omega_{i_1,..,i_{k_A}}^{2q}  +o(\vert J \vert^{2q})
\nonumber \\
&&\propto \sum_{n=0}^{L_A-1} \sum_{m=0}^{L_B-1} \Delta^{2q}(1+n+m) \frac{2^{n+m}}{\sqrt {n+m} }
\nonumber \\
&&\propto \sum_{n=0}^{L_A-1} \sum_{m=0}^{L_B-1} \left[  
\frac{v}{(n+m)^a} 2^{-b (n+m)} 
 \right]^{2q}  \frac{2^{n+m}}{\sqrt {n+m} }
\nonumber \\
&&\propto v^{2q} \sum_{n=0}^{L_A-1} \sum_{m=0}^{L_B-1}  \frac{2^{(n+m)(1-2qb)}}{(n+m)^{\left(\frac{1}{2}+2qa  \right)} }
\label{entqavqsmallres}
\end{eqnarray}
For the bipartite case $L_A=L_B=L$, this yields the following power-laws
in terms of the size ${\cal M}=2^{2L}$ of the Hilbert space
\begin{eqnarray}
\overline{ S_q } 
&&\propto v^{2q} {\cal M}^{-\tau_b(q)}
\label{entqavqsmallrestau}
\end{eqnarray}
with the multifractal exponents
\begin{eqnarray}
\tau_b(q) && = 2qb-1  \ \ \ \ {\rm for } \ \ \ q \leq \frac{1}{2b}
\nonumber \\
\tau_b(q) && =0  \ \ \ \ \ \ \ \ {\rm for } \ \ \  \frac{1}{2b} \leq q \leq \frac{1}{2}
\label{taubq}
\end{eqnarray}

In the region $q>\frac{1}{2} $, Eq. \ref{entqavqlarge} yields
\begin{eqnarray}
\tau_b(q) && =0  \ \ \ \ {\rm for } \ \ \ q \geq \frac{1}{2}
\label{taubqbis}
\end{eqnarray}
in continuity with the result of Eq. \ref{taubq}.

\subsection{ Multifractal spectrum $f^{criti}(\alpha)$ at the critical point $b_c=1$  }

At the critical point $b_c=1$, the multifractal exponents of Eqs \ref{taubq}
and \ref{taubqbis}
\begin{eqnarray}
\tau^{criti}(q) && = 2q-1  \ \ \ \ {\rm for } \ \ \ q \leq \frac{1}{2}
\nonumber \\
\tau^{criti}(q) && =0  \ \ \ \ \ \ \ \ \ \ \ \ \ {\rm for } \ \ \  \frac{1}{2} \leq q 
\label{tauqcriti}
\end{eqnarray}
correspond to the Legendre transform of the multifractal spectrum
\begin{eqnarray}
f^{criti}(\alpha) = \frac{\alpha}{2} \ \ \ {\rm for} \ \  0 \leq \alpha \leq 2
\label{fstrong}
\end{eqnarray}
that is well-known as the
'strong-multifractality' spectrum 
in the context of Anderson Localization Transition
in the limit of infinite dimension $d \to +\infty$ \cite{mirlinrevue}
or in the long-ranged power-law Anderson model in one-dimension of Eq. \ref{prbm} 
 \cite{levitov1,levitov2,levitov3,levitov4,mirlin_evers,fyodorov,fyodorovrigorous,oleg1,oleg2,oleg3,oleg4,olivier_per,olivier_strong,olivier_conjecture,us_strongmultif}.

More recently, the strong multifractality spectrum of Eq. \ref{fstrong}
has been found to describe the statistics of matrix elements of local operators
and the statistics of hybridization Ratios in {\it nearest-neighbors } Many-Body-Localized models at criticality \cite{c_mblstrongmultif}. Our conclusion is thus that the present toy model is not so far from more realistic short-ranged models, since it is able to reproduce the same critical multifractal spectrum.

\subsection{ Multifractal spectrum in the Many-Body-Localized Phase $b>b_c=1$  }

In the many-Body-Localized phase $b>b_c=1$, 
 the multifractal exponents of Eqs \ref{taubq}
and \ref{taubqbis}
\begin{eqnarray}
\tau^{loc}_b(q) && =  2qb-1  \ \ \ \ {\rm for } \ \ \ q \leq \frac{1}{2b}
\nonumber \\
\tau^{loc}_b(q) && =0 \ \ \ \ \ \ \ \ \ \ \ \ {\rm for } \ \ \ \frac{1}{2b} \leq q
\label{tauqloc}
\end{eqnarray}
is the Legendre transform of the multifractal spectrum
\begin{eqnarray}
f^{loc}_b(\alpha) = \frac{\alpha}{2 b} \ \ \ {\rm for} \ \  0 \leq \alpha \leq 2 b
\label{fstrongloc}
\end{eqnarray}
that is a very simple deformation of the critical result of Eq. \ref{fstrong}.
The fact that multifractality occurs also in the Many-Body-Localized Phase
is in agreement with the analysis of matrix elements of local operators
and the statistics of hybridization Ratios in {\it nearest-neighbors } Many-Body-Localized models \cite{c_mblstrongmultif}.

\section { Conclusions }

\label{sec_conclusion}

In this paper, in analogy with the strong disorder limit of the Anderson transition for a single particle (recalled in \ref{anderson}), we have considered 
the strong disorder limit of the MBL transition, defined as the limit
where the level statistics at the MBL critical point is close to the Poisson statistics of the MBL phase.
 For a quantum one-dimensional toy model, we have analyzed the statistical properties of the rare extensive resonances that are needed to destabilize the Many-Body Localized phase. At criticality, we have found that the entanglement entropy can grow with an exponent $0<\alpha=3/2-a < 1$ anywhere between the area law $\alpha=0$ and the volume law $\alpha=1$, as a function of the power-law exponent $a$ of 
the couplings (Eq. \ref{jr}), while the entanglement spectrum follows the strong multifractality statistics of Eq. \ref{fstrong}, well-known as the
'strong-multifractality' spectrum 
in the context of Anderson Localization Transition \cite{mirlinrevue},
and found recently for {\it nearest-neighbor } MBL models at criticality \cite{c_mblstrongmultif}.
 
For an initial short-ranged model, we thus expect that the important extended rare resonances are described by some effective renormalized couplings $J_{i_1,...,i_k}$ and it would be interesting to compute their properties in terms of the initial parameters, either via the forward approximation \cite{levitov,luca,emergent_ros,forward,qrem} or via some RG procedure.  

The main difference between the present approach and the existing RG on resonances \cite{vasseur_resonant} is as follows : 

(i) here, in the strong disorder limit, we have considered the resonances concerning a single eigenstate in the middle of the spectrum. 

(ii) on the contrary, Ref. \cite{vasseur_resonant}
 is based on the notion of resonant blocks,
so that when two blocks are declared to be resonant, all eigenstates of the two blocks are strongly mixed and exhibit level repulsion.
 So we feel that this assumption should be valid
in the opposite weak-disorder limit of the transition, defined as the limit 
where the level statistics at the critical point is close
to the Wigner-Dyson statistics of the delocalized phase.

So we believe that (i) and (ii) are actually
 the two extreme theories of a continuous family of MBL critical points,
where the level statistics interpolates between the Poisson statistics
 and the Wigner-Dyson statistics.

 Finally, in the MBL phase near criticality, we have obtained the simple value $\nu=1$ for the correlation length exponent, that simply reflects the crossing of the exponential decay of the level spacing and of the exponential decay of the couplings.
 More generally, and independently of the strong disorder limit, we have explained why the correlation length exponent $\nu$ of the MBL transition has no reason to satisfy the usual Harris inequality $\nu \geq 2/d$, so that there is actually no theoretical inconsistency in the numerical works where the correlation length exponent $\nu$ is found to violate the Harris inequality \cite{kjall,alet}.

\section*{ Acknowledgments }

It is a pleasure to thank Romain Vasseur for an illuminating discussion
on various aspects of the MBL transition, 
and in particular for his explanations 
concerning the RG procedure on resonances \cite{vasseur_resonant}.

\appendix

\section{ Off-diagonal elements of the reduced density matrix  }

\label{app_off}

In this Appendix, we explain why the off-diagonal elements 
of Eq. \ref{Raaprime}
of the reduced density matrix of Eq. \ref{rhoa} can be neglected 
at first order in the couplings.

Let us consider the probability distribution
$ P_{i_1,..,i_{k_A}; i_1',..,i_{k_A'}'}(R (i_1,..,i_{k_A} ; i_1',..,i_{k_A'}' ))$
of the variable $ R (i_1,..,i_{k_A} ; i_1',..,i_{k_A'}') $ of Eq. \ref{Raaprime} when the indices $(i_1,..,i_{k_A};i_1',..,i_{k_A'}')$
are all distincts (when some indices are the same, the generalization is straighforward).
The Fourier transform at lowest order in the couplings reads
\begin{eqnarray}
 \overline{ e^{ i t  R (i_1,..,i_{k_A} ; i_1',..,i_{k_A'}')  }  } && = \int_{-\infty} ^{+\infty} d R (i_1,..,i_{k_A} ; i_1',..,i_{k_A'}') 
P_{i_1,..,i_{k_A}; i_1',..,i_{k_A'}'}(R (i_1,..,i_{k_A} ; i_1',..,i_{k_A'}' ))
e^{ i t  R (i_1,..,i_{k_A} ; i_1',..,i_{k_A'}')  }
\nonumber \\
&& =\overline{ e^{ it   \sum_{k_B=1}^{L_B}
 \sum_{L_A+1 \leq j_1< j_2..<j_{k_B} \leq {L_A+L_B}}
\left(\frac{ J_{i_1,..,i_{k_A},j_1,...j_{k_B}} }{ 2 \sum_{q=1}^{k_A} h_{i_q}+2 \sum_{p=1}^{k_B} h_{j_p}} \right)
\left(\frac{ J_{i_1',..,i_{k_A}',j_1,...j_{k_B}} }{ 2 \sum_{q=1}^{k_A'} h_{i_q'}
+2 \sum_{p=1}^{k_B} h_{j_p} } \right) }}
\nonumber \\
&& =\overline{ 
\prod_{k_B=1}^{L_B}
 \prod_{L_A+1 \leq j_1< j_2..<j_{k_B} \leq {L_A+L_B}}
\left[1- \left(1- e^{i t \left(\frac{ J_{i_1,..,i_{k_A},j_1,...j_{k_B}} }{ 2 \sum_{q=1}^{k_A} h_{i_q}+2 \sum_{p=1}^{k_B} h_{j_p}} \right)
\left(\frac{ J_{i_1',..,i_{k_A}',j_1,...j_{k_B}} }{ 2 \sum_{q=1}^{k_A'} h_{i_q'}
+2 \sum_{p=1}^{k_B} h_{j_p} }  \right) }  \right)  \right] }
\nonumber \\
&& = 
1 -  \sum_{k_B=1}^{L_B}
 \sum_{L_A+1 \leq j_1< j_2..<j_{k_B} \leq {L_A+L_B}}
\overline{ \left(1-e^{i t \left(\frac{ J_{i_1,..,i_{k_A},j_1,...j_{k_B}} }{ 2 \sum_{q=1}^{k_A} h_{i_q}+2 \sum_{p=1}^{k_B} h_{j_p}} \right)
\left(\frac{ J_{i_1',..,i_{k_A}',j_1,...j_{k_B}} }{ 2 \sum_{q=1}^{k_A'} h_{i_q'}
+2 \sum_{p=1}^{k_B} h_{j_p} }  \right) } \right) }
  +..
\label{fourier}
\end{eqnarray}

The variables 
\begin{eqnarray}
E_A && =2 \sum_{q=1}^{k_A} h_{i_q}
\nonumber \\
E_A' && =2 \sum_{q=1}^{k_A'} h_{i_q'}
\nonumber \\
E_ B && = 2 \sum_{p=1}^{k_B} h_{j_p}
\label{esumab}
\end{eqnarray}
are three independent Gaussian variables (Eq. \ref{gauss})
of zero mean and of variances $4 W^2 k_A$, $4 W^2 k_A'$ 
and $4 W^2 k_B$ respectively.
As a consequence, the probability distribution $Q(z)$ of
the variable $z=(E_A+E_B)(E_A'+E_B)  $ involved in Eq. \ref{fourier}
reads using the change of variables $y=E_A+E_B$ and $y'=E_A+E_B $

\begin{eqnarray}
&& Q(z)  = \int_{-\infty}^{+\infty} \frac{dE_B e^{- \frac{ E_B^2}{ 8 k_B W^2} }}{2 \sqrt { 2 k_B  \pi W^2 }}
\int_{-\infty}^{+\infty} \frac{dE_A e^{- \frac{ E_A^2}{ 8 k_A W^2} }}{2 \sqrt { 2 k_A  \pi W^2 }}
\int_{-\infty}^{+\infty} \frac{dE_A' e^{- \frac{ (E_A')^2}{ 8 k_A' W^2} }}{2 \sqrt { 2 k_A'  \pi W^2 }}
\delta \big(z- (E_A+E_B)(E_A'+E_B) \big)
\nonumber \\
&& = \int_{-\infty}^{+\infty} \frac{dE_B e^{- \frac{ E_B^2}{ 8 k_B W^2} }}{2 \sqrt { 2 k_B  \pi W^2 }}
\int_{-\infty}^{+\infty} \frac{dy e^{- \frac{ (y-E_B)^2}{ 8 k_A W^2} }}{2 \sqrt { 2 k_A  \pi W^2 }}
\int_{-\infty}^{+\infty} \frac{dy' e^{- \frac{ (y'-E_B)^2}{ 8 k_A' W^2} }}{2 \sqrt { 2 k_A'  \pi W^2 }}
\delta \big(z- y y' \big)
\nonumber \\
&& = 
\int_{-\infty}^{+\infty} \frac{dy e^{- \frac{ y^2}{ 8 k_A W^2} }}{ \sqrt { 8 k_A  \pi W^2 }}
\int_{-\infty}^{+\infty} \frac{dy' e^{- \frac{(y')^2 }{ 8 k_A' W^2} }}{ \sqrt { 8 k_A'  \pi W^2 }}
\frac{ \delta \big( y'-\frac{z}{y} \big) }{ \vert y \vert} 
\int_{-\infty}^{+\infty} \frac{dE_B}{ \sqrt { 8 k_B  \pi W^2 }}
 e^{- \frac{ E_B^2}{ 8  W^2} \left( \frac{1}{k_B} +\frac{1}{k_A}+\frac{1}{k_A'}\right) }
 e^{ \frac{  (y+y') E_B}{ 4 k_A W^2} }
\nonumber \\
&& =  e^{ \frac{  z  }{ 4  W^2  \left( k_A+k_A'+ \frac{k_A k_A'}{k_B}  \right)} }
\frac{1}{ 4 \pi W^2 \sqrt {  k_A k_A'+k_B(k_A+k_A') }}
K_0 \left( \vert z \vert \frac{ \sqrt{(k_B+k_A)(k_B+k_A')} }{ 4 W^2\left( k_A+k_A'+ \frac{k_A k_A'}{k_B}  \right) } \right)
\label{gaussprod}
\end{eqnarray}
in terms of the Bessel function $K_0(u)$ displaying the logarithmic singularity near the origin
\begin{eqnarray}
K_0(u) = - \ln u +(\ln 2-\gamma_{Euler})+O(u^2)
\label{bessel}
\end{eqnarray}
and an exponential decay for large $u$.

As a consequence, the average of the inverse of the variable $z$ remains finite
\begin{eqnarray}
 \int_{-\infty} ^{+\infty} \frac{dz}{z}  Q(z) && = \int_0^{+\infty} \frac{dz}{z} \left[ Q(z)-Q(-z)  \right]
\nonumber \\
&& =  \int_0^{+\infty} \frac{dz}{z} 
\frac{ \sinh \left( \frac{  z  }{ 4  W^2  \left( k_A+k_A'+ \frac{k_A k_A'}{k_B}  \right)} \right)}{ 2 \pi W^2 \sqrt {  k_A k_A'+k_B(k_A+k_A') }}
K_0 \left( \vert z \vert \frac{ \sqrt{(k_B+k_A)(k_B+k_A')} }{ 4 W^2\left( k_A+k_A'+ \frac{k_A k_A'}{k_B}  \right) } \right)
\nonumber \\
&& = 
\frac{1}{ 4 W^2 \sqrt {  k_A k_A'+k_B(k_A+k_A') }}
\arcsin \left( \frac{1}{(k_B+k_A)(k_B+k_A')}  \right) 
\label{avinverse}
\end{eqnarray}

So Eq. \ref{fourier} yields that the variable $ R (i_1,..,i_{k_A} ; i_1',..,i_{k_A'}') $ has the finite average value
\begin{eqnarray}
 \overline{   R (i_1,..,i_{k_A} ; i_1',..,i_{k_A'}')    } && =
  \sum_{k_B=1}^{L_B}
 \sum_{L_A+1 \leq j_1< .<j_{k_B} \leq {L_A+L_B}}
\frac{J_{i_1,..,i_{k_A},j_1,...j_{k_B}} J_{i_1',..,i_{k_A}',j_1,...j_{k_B}}}
{ 4 W^2 \sqrt {  k_A k_A'+k_B(k_A+k_A') }}
\arcsin \left( \frac{1}{(k_B+k_A)(k_B+k_A')}  \right) 
\nonumber \\
&& +o(J^2)
\label{fourierR}
\end{eqnarray}
which is of second order $O(J^2)$ with respect to the couplings.
As a consequence, at the first order in the couplings considered
in the main text, the off-diagonal elements are negligible.

\section{ Disorder-averaged values $\overline{Y_q}$ }

\label{app_yqav}

Using the identity 
\begin{eqnarray}
\frac{1}{a^q} = \frac{1}{\Gamma(q)} \int_0^{+\infty} dt \  t^{q-1} e^{-at} 
\label{defgamma}
\end{eqnarray}
the disorder-averaged value of Eq. \ref{ipr} 
can be decomposed into the two contributions
\begin{eqnarray}
\overline{Y_q} && = 
\overline{ \frac{ 1 +\Sigma_q  }{ (1+\Sigma_1)^q} }
=  \overline{ Y_q }\vert_{first} 
+ \overline{ Y_q }\vert_{second} 
\nonumber \\
 \overline{ Y_q }\vert_{first} && =\overline{ \frac{ 1  }{ (1+\Sigma_1)^q} } =
\frac{1}{\Gamma(q)} \int_0^{+\infty} dt \ t^{q-1} e^{-t}
\ \overline{  e^{-t \Sigma_1 }}
\nonumber \\
 \overline{ Y_q }\vert_{second} && =\overline{ \frac{ \Sigma_q  }{ (1+\Sigma_1)^q} } =
\frac{1}{\Gamma(q)} \int_0^{+\infty} dt \ t^{q-1} e^{-t} \ 
\overline{ \Sigma_q   e^{-t \Sigma_1 }}
\label{iprav}
\end{eqnarray}

Eq. \ref{lapsigma1d} yields that the first contribution reads
\begin{eqnarray}
 \overline{ Y_q }\vert_{first} && = 
\frac{1}{\Gamma(q)} \int_0^{+\infty} dt \ t^{q-1} e^{-t}
(1-  \Omega_1  t^{\frac{1}{2}} +o(J) ) 
\nonumber \\
&& = 1 -  \Omega_1  \frac{\Gamma \left(q+ \frac{1}{2}\right)}{\Gamma(q)}+o(J)
\label{yqavfirst}
\end{eqnarray}
To evaluate the second contribution, one needs to consider
separately
the two cases $0<q<\frac{1}{2} $ and $q>\frac{1}{2} $.

\subsection{ Region $0<q<\frac{1}{2} $ }

 For $0<q<\frac{1}{2} $, the average of $\Sigma_q $ is finite 
(Eq. \ref{avsigmaq}) so that the second contribution at leading order
\begin{eqnarray}
 \overline{ Y_{q<1/2} }\vert_{second} && =\overline{  \Sigma_q  }
= \frac{ \Gamma \left(
\frac{1}{2}-q  \right) }{4^q \sqrt{\pi}}
\sum_{k_A=1}^{L} 
 \sum_{1 \leq i_1< i_2..<i_{k_A} \leq {L}} \Omega_{i_1,..,i_{k_A}}^{2q} 
 +o(\vert J \vert^{2q})
\label{yqavsecondqsmallerdemi}
\end{eqnarray}
is bigger than the correction of order $\Omega_1=O(J)$ 
of the first contribution of Eq. \ref{yqavfirst}, so that
the sum of the two contributions reads
\begin{eqnarray}
 \overline{ Y_{q<1/2} } && =1+ \frac{ \Gamma \left(
\frac{1}{2}-q  \right) }{4^q \sqrt{\pi}}
\sum_{k_A=1}^{L} 
 \sum_{1 \leq i_1< i_2..<i_{k_A} \leq {L}}\Omega_{i_1,..,i_{k_A}}^{2q} 
 +o(\vert J \vert^{2q})
\label{yqavqsmallerdemi}
\end{eqnarray}

This disorder-averaged value thus coincides at leading order
with the typical value computed in Eq. \ref{entqavqsmall}
\begin{eqnarray}
Y_{q<1/2}^{typ}  \equiv e^{ \overline{ \ln Y_{q<1/2} } }
= e^{  (1-q) \overline{S_q } }
 = 1+\frac{ \Gamma \left(
\frac{1}{2}-q  \right) }{4^q \sqrt{\pi}} \sum_{k_A=1}^{L} 
 \sum_{1 \leq i_1< i_2..<i_{k_A} \leq {L}} \Omega_{i_1,..,i_{k_A}}^{2q} 
 +o(\vert J \vert^{2q})
\label{yqtypqsmallerdemi}
\end{eqnarray}

\subsection{ Region $q>\frac{1}{2} $ }

For $q >\frac{1}{2} $, the average of $\Sigma_q $ is infinite,
so one needs to evaluate the divergence for small $t$ of
\begin{eqnarray}
 \overline{ \Sigma_q   e^{-t \Sigma_1 }}
&& =  \overline{ \sum_{k_A=1}^{L_A}  \sum_{1 \leq i_1< i_2..<i_{k_A} \leq {L_A}}
D^q_{i_1,..,i_{k_A}}   e^{-t \sum_{k_A'=1}^{L_A}  \sum_{1 \leq i_1'< i_2'..<i_{k_A}' \leq {L_A}} D_{i_1',..,i_{k_A}'}}}
\nonumber \\
&& =  \sum_{k_A=1}^{L_A}  \sum_{1 \leq i_1< i_2..<i_{k_A} \leq {L_A}}
\overline{ D^q_{i_1,..,i_{k_A}}   e^{-t  D_{i_1,..,i_{k_A}}} } \left[1 + O(t^{\frac{1}{2}}) \right]
\label{sigqsig1}
\end{eqnarray}

The L\'evy distribution of Eqs \ref{levy}, \ref{levylaplace} for the variable $ D_{i_1,..,i_{k_A}}$ yields the singularity
\begin{eqnarray}
\overline{ D^q_{i_1,..,i_{k_A}}   e^{-t  D_{i_1,..,i_{k_A}}} } &&
= \frac{\Omega_{i_1,..,i_{k_A}}}{2 \sqrt{\pi}  }
\int_0^{+\infty} dD D^{q-\frac{3}{2}} e^{- \frac{\Omega_{i_1,..,i_{k_A}}^2}{4 D} } e^{-t D}
\nonumber \\
&& =\frac{\Omega_{i_1,..,i_{k_A}}}{2 \sqrt{\pi}  } t^{\frac{1}{2}-q}
\int_0^{+\infty} dx x^{q-\frac{3}{2}} e^{- \frac{\Omega_{i_1,..,i_{k_A}}^2}{4 x} t } e^{- x }
\nonumber \\
&& =\frac{\Omega_{i_1,..,i_{k_A}}}{2 \sqrt{\pi}  } t^{\frac{1}{2}-q}
\Gamma \left( q- \frac{1}{2}\right) +o(\vert J \vert)
\label{auxidq}
\end{eqnarray}
So Eq. \ref{sigqsig1} displays the same singularity in terms of the parameter of Eq. \ref{omega1}
\begin{eqnarray}
 \overline{ \Sigma_q   e^{-t \Sigma_1 }}
&& = \frac{\Omega_1}{2 \sqrt{\pi}  } t^{\frac{1}{2}-q}
\Gamma \left( q- \frac{1}{2}\right) +o(\vert J \vert) 
\label{sigqsig1dv}
\end{eqnarray}
and the second contribution of Eq. \ref{iprav}
reads at leading order 
\begin{eqnarray}
 \overline{ Y_{q>\frac{1}{2} } }\vert_{second} &&  =
\frac{1}{\Gamma(q)} \int_0^{+\infty} dt \ t^{q-1} e^{-t} \ 
\overline{ \Sigma_q   e^{-t \Sigma_1 }}
\nonumber \\
&& =\frac{\Omega_1}{2   }
 \frac{\Gamma \left( q- \frac{1}{2}\right)}{\Gamma(q)} 
 +o(\vert J \vert) 
\label{iprsecondqbigger}
\end{eqnarray}
The sum with the first contribution of Eq. \ref{yqavfirst}
yields the disorder-average of $Y_q$ in the region $q>1/2$
\begin{eqnarray}
 \overline{ Y_{q>\frac{1}{2} } } 
&& = 1 
+ \frac{\Omega_1}{2 \Gamma(q)  }
\left[ \Gamma \left( q- \frac{1}{2}\right) - 2 \Gamma \left(q+ \frac{1}{2}\right) \right] 
+o(J)\nonumber \\
&& = 1- \Omega_1 \frac{\Gamma \left( q- \frac{1}{2}\right)}{\Gamma(q-1)} 
+o(J)
\label{yqavqbig}
\end{eqnarray}

This disorder-averaged value is thus different
from the typical value computed in Eq. \ref{entqavqlarge}
\begin{eqnarray}
Y_{q>1/2}^{typ}  \equiv e^{ \overline{ \ln Y_{q>1/2} } }
= e^{  (1-q) \overline{S_q } }
 = 1+  \Omega_1 \sqrt{\pi}\left[
 \frac{ 1 }{  \sin \left(\frac{\pi}{2q} \right)} -q \right]+o(J)
\label{yqtypqbiggerdemi}
\end{eqnarray}

\subsection{ Consequence for the variance of the entanglement entropy }

In the vicinity of $q=1$, the difference found above between
the averaged value of Eq. \ref{yqavqbig}
\begin{eqnarray}
 \overline{ Y_{q>\frac{1}{2} } } 
 =  \overline{ e^{(q-1)S_q} } = 1+ (q-1) \overline{ S_q} + \frac{(q-1)^2}{2} 
\overline{ S_q^2} +O(q-1)^3
\label{yqavqbigexpand}
\end{eqnarray}
and the typical value of Eq. \ref{yqtypqbiggerdemi}
\begin{eqnarray}
Y_{q>1/2}^{typ}  
= e^{  (1-q) \overline{S_q } }
 = 1+ (q-1) \overline{ S_q} + \frac{(q-1)^2}{2} 
(\overline{ S_q})^2 +O(q-1)^3
\label{yqtypqbiggerdemiexpand}
\end{eqnarray}
yields the variance of the entanglement entropy $S_1$
\begin{eqnarray}
\overline{ S_1^2} - (\overline{ S_1})^2
&& =2 \lim \limits_{q \to 1} \left[  \frac{ \overline{ Y_{q>\frac{1}{2} } } - Y_{q>1/2}^{typ} } {(q-1)^2 } \right]
\nonumber \\
&&    =2 \Omega_1 \lim \limits_{q \to 1} \left[  \frac{
 \sqrt{\pi}\left[ q- \frac{ 1 }{  \sin \left(\frac{\pi}{2q} \right)}  \right]
-  \frac{\Gamma \left( q- \frac{1}{2}\right)}{\Gamma(q-1)}  } {(q-1)^2 } \right] + o(J)
\nonumber \\
&&    = \Omega_1 \sqrt{\pi} \left( 4 \ln 2 - \frac{\pi^2}{4}  \right) + o(J)
\label{entvar}
\end{eqnarray}
So the scale $\Omega_1$ that governs the scale of the disorder-averaged
entanglement entropy (Eq. \ref{entavq1}) also determines its variance
(Eq. \ref{entvar}).


\begin{thebibliography}{99}

\bibitem{deutsch}
J.M. Deutsch, Phys. Rev. A 43, 2046 (1991).

\bibitem{srednicki}
M. Srednicki, Phys. Rev. E 50, 888 (1994).

\bibitem{nature}
M. Rigol, V. Dunjko and M. Olshanii, Nature 452, 854 (2008)

\bibitem{mite}
S. Goldstein, D.A. Huse, J.L. Lebowitz and R. Tumulka, Phys. Rev. Lett. 115,
100402 (2015).

\bibitem{rigol}
L. D'Alessio, Y. Kafri, A. Polkovnikov and M. Rigol, arxiv:1509.06411.

\bibitem{revue_huse}
R. Nandkishore and D. A. Huse, Ann. Review of Cond. Mat. Phys. 6, 15 (2015).

\bibitem{revue_altman}
 E. Altman and R. Vosk, Ann. Review of Cond. Mat. Phys. 6, 383 (2015).


\bibitem{bauer}
B. Bauer and C. Nayak, J. Stat. Mech. P09005 (2013).



\bibitem{pekker1}
D. Pekker and B.K. Clark, arxiv:1410.2224.

\bibitem{pekker2}
X. Yu, D. Pekker and B.K. Clark, arxiv:1509.01244.

\bibitem{friesdorf}
M. Friesdorf, A.H. Werner, W. Brown, V. B. Scholz and J. Eisert, Phys. Rev. Lett.
114, 170505 (2015).

\bibitem{sondhi}
V. Khemani, F. Pollmann and S. L. Sondhi, arxiv:1509.00483.

\bibitem{tensor}
A. Chandran, J. Carrasquilla, I.H. Kim, D.A. Abanin and G. Vidal, Phys. Rev. B 92,
024201 (2015).


\bibitem{fisher_AF}
D. S. Fisher, Phys. Rev. B 50, 3799 (1994).

\bibitem{fisher}
D. S. Fisher, Phys. Rev. Lett. 69, 534 (1992); \\
D. S. Fisher, Phys. Rev. B 51, 6411 (1995).

\bibitem{fisherreview}
D. S. Fisher, Physica A 263, 222 (1999).



\bibitem{vosk_dyn1}
R. Vosk and E. Altman, Phys. Rev. Lett. 110, 067204 (2013).

\bibitem{vosk_dyn2}
R. Vosk and E. Altman, Phys. Rev. Lett. 112, 217204 (2014).



\bibitem{rsrgx}
D. Pekker, G. Refael, E. Altman, E. Demler and V. Oganesyan,
Phys. Rev. X 4, 011052 (2014).

 \bibitem{rsrgx_moore}
Y. Huang and J.E. Moore, Phys. Rev. B 90, 220202(R) (2014).

\bibitem{vasseur_rsrgx}
R. Vasseur, A. C. Potter and S.A. Parameswaran, Phys. Rev. Lett. 114, 217201 (2015).


\bibitem{yang_rsrgx}
M. Pouranvari and K. Yang, Phys. Rev. B 92, 245134 (2015).

\bibitem{rsrgx_bifurcation}
 Y.Z. You, X.L. Qi and C. Xu, arxiv:1508.036035


\bibitem{c_emergent}
C. Monthus, J. Stat. Mech. 033101 (2016).


\bibitem{emergent_swingle}
B. Swingle, arxiv:1307.0507.

\bibitem{emergent_serbyn}
M. Serbyn, Z. Papic and D.A. Abanin, Phys. Rev. Lett. 111, 127201 (2013).

\bibitem{emergent_huse}
D.A. Huse, R. Nandkishore and V. Oganesyan, Phys. Rev. B 90, 174202 (2014).

\bibitem{emergent_ent}
A. Nanduri, H. Kim and D.A. Huse, Phys. Rev. B 90, 064201 (2014).

\bibitem{imbrie}
J. Z. Imbrie, arxiv:1403.7837.

\bibitem{serbyn_quench}
M. Serbyn, Z. Papic and D.A. Abanin, Phys. Rev. B 90, 174302 (2014).

\bibitem{emergent_vidal}
A. Chandran, I.H. Kim, G. Vidal and  D.A. Abanin, Phys. Rev. B 91, 085425 (2015).

\bibitem{emergent_ros}
V. Ros, M. M\"uller and A. Scardicchio, Nucl. Phys. B 891, 420 (2015).


\bibitem{vosk_rgentanglement}
R. Vosk, D.A. Huse, and E. Altman, Phys. Rev. X 5, 031032 (2015).

\bibitem{vasseur_resonant}
A. C. Potter, R. Vasseur and S.A. Parameswaran, Phys. Rev. X 5, 031033 (2015).


\bibitem{grover}
T. Grover, arxiv:1405.1471.

\bibitem{harrisMBL}
A. Chandran, C.R. laumann and V. Oganesyan, arxiv:1509.04285.

\bibitem{kjall}
J. A. Kj\"all, J. H. Bardarson and F. Pollmann, Phys. Rev. Lett. 113, 107204 (2014).

\bibitem{alet}
D. J. Luitz, N. Laflorencie and F. Alet, Phys. Rev. B 91, 081103 (2015).

\bibitem{levitov}
B.L. Altshuler, Y. Gefen, A. Kamenev and L.S. Levitov,
Phys. Rev. Lett. 78, 2803 (1997).

\bibitem{gornyi_fock}
I.V. Gornyi, A.D. Mirlin and D.G. Polyakov, Phys. Rev. Lett. 95, 206603 (2005).


\bibitem{vadim}
V. Oganesyan and D.A. Huse, Phys. Rev. B 75, 155111 (2007).


\bibitem{us_mblaoki}
C. Monthus and T. Garel, Phys. Rev. B 81, 134202 (2010).

\bibitem{luca}
A. De Luca, B.L. Altshuler, V.E. Kravtsov and A. Scardicchio, Phys. Rev. Lett. 113, 046806 (2014).

\bibitem{gornyi}
I.V. Gornyi, A.D. Mirlin, and D.G. Polyakov, Phys. Rev. B 93, 125419 (2016).

\bibitem{c_mblstrongmultif}
C. Monthus, arxiv:1603.04701.

\bibitem{mirlinrevue}
 F. Evers and A.D. Mirlin, Rev. Mod. Phys. 80, 1355 (2008).

\bibitem{levitov1}
L.S. Levitov, Europhys. Lett. 9, 83 (1989).

\bibitem{levitov2}
L.S. Levitov, Phys. Rev. Lett. 64, 547 (1990).

\bibitem{levitov3}
B.L. Altshuler and L.S. Levitov, Phys. Rep. 288, 487 (1997).

\bibitem{levitov4}
L.S. Levitov, Ann. Phys. (Leipzig) 8, 5, 507 (1999).

\bibitem{mirlin_evers}
 F. Evers and A. D. Mirlin
Phys. Rev. Lett. 84, 3690 (2000); \\
A.D. Mirlin and F. Evers, 
Phys. Rev. B 62,  7920 (2000).

\bibitem{fyodorov}
Y.V. Fyodorov, A. Ossipov and A. Rodriguez, J. Stat. Mech. L12001 (2009).

\bibitem{fyodorovrigorous}
Y.V. Fyodorov, A. Kupiainen and C. Webb. arxiv:1509.01366.


\bibitem{oleg1}
O. Yevtushenko and V. E. Kratsov,
J. Phys. A 36, 8265 (2003).

\bibitem{oleg2}
O. Yevtushenko and A. Ossipov, 
J. Phys. A 40, 4691 (2007).

\bibitem{oleg3}
S. Kronm\"uller, O. M.  Yevtushenko and E. Cuevas, 
J. Phys. A 43, 075001 (2010).

\bibitem{oleg4}
V. E. Kratsov, A. Ossipov, O. M.  Yevtushenko and E. Cuevas, 
Phys. Rev. B 82, 161102(R) (2010).

\bibitem{olivier_per}
E. Bogomolny and O. Giraud, Phys. Rev. E 84, 036212 (2012).

\bibitem{olivier_strong}
E. Bogomolny and O. Giraud, Phys. Rev. E 84, 046208 (2012).

\bibitem{olivier_conjecture}
E. Bogomolny and O. Giraud, Phys. Rev. Lett. 106, 044101 (2011).


\bibitem{us_strongmultif}
C. Monthus and T. Garel, J. Stat. Mech. (2010) P09015.

\bibitem{levy}
P. L\'evy, ``Th\'eorie de l'addition des variables al\'eatoires'',
Gauthier-Villars, Paris (1937).

\bibitem{jpbreview}
J.P. Bouchaud and A. Georges, Phys. Rep. 195 , 127 (1990).

\bibitem{Der}
B. Derrida, Physica D 107, 186 (1997).

\bibitem{Der_Fly}
B. Derrida and H. Flyvbjerg, J. Phys. A Math. Gen. 20, 5273 (1987).

 \bibitem{us_critiweights}
C. Monthus and T. Garel, Phys. Rev. E 75, 051119 (2007).

\bibitem{harris}
A.B. Harris, J. Phys. C 7, 1671 (1974).

\bibitem{chayes}
J.T. Chayes, L. Chayes, D.S. Fisher and T. Spencer,
 Phys. Rev. Lett. 57, 2999
(1986).  


\bibitem{kunz}
H. Kunz and B. Souillard, J. Phys. Lett. 44, L411 (1983).

\bibitem{mirlin_fyodorov}
A.D. Mirlin and Y.V. Fyodorov, Nucl. Phys. B 366, 507 (1991).


\bibitem{us_andersontreeTW}
C. Monthus and T. Garel, J. Phys. A Math. Theor. 42, 075002 (2009).

\bibitem{forward}
F. Pietracaprina, V. Ros, A. Scardicchio, Phys. Rev. B 93, 054201 (2016).

\bibitem{qrem}
C.R. Laumann, A. Pal and A. Scardicchio, Phys. Rev. Lett. 113, 200405 (2014); \\
C.L. Baldwin, C.R. Laumann, A. Pal and A. Scardicchio, arxiv:1509.08926.


\bibitem{luca_mbl}
A. de Luca and A. Scardicchio, EPL 101, 37003 (2013).

\bibitem{santos}
E.J. Torres-Herrera and L.S. Santos, Phys. Rev. B 92, 014208 (2015).

\bibitem{fradkin}
X. Chen {\it et al.}, Phys. Rev. B 92, 214204 (2015).



\bibitem{serbyn}
M. Serbyn, Z. Papic and D.A. Abanin, Phys. Rev. X 041047 (2015).

\bibitem{mbl_dyson}
M. Serbyn and J.E. Moore, Phys. Rev. B 93, 041424 (2016).

\bibitem{c_mbldysonbrownian}
C. Monthus, J. Stat. Mech. 033113 (2016).

 \end{thebibliography}
\end{document}